\documentclass[a4paper,11pt]{article}

\usepackage{jheppub}

\usepackage[T1]{fontenc} 
\usepackage[final]{showlabels} 
\usepackage{amsmath}  
\usepackage{amssymb}  
\usepackage{latexsym}
\usepackage{enumitem} 
\usepackage{lipsum}
\usepackage{graphicx}
\usepackage{subfig} 
\usepackage{empheq}
\usepackage{upgreek} 
\usepackage{cancel}
\usepackage{csquotes}
\usepackage{ytableau}
\usepackage{hyperref}
\usepackage{pifont}
\usepackage{csquotes}
\usepackage{ytableau}
\usepackage{blkarray}
\usepackage{multirow}
\usepackage{lscape}
\usepackage{float}

\usepackage{braket}
\usepackage{amssymb}
\usepackage{amsmath}
\usepackage{amsthm}
\usepackage{mathrsfs}
\usepackage{skak}
\usepackage{framed}
\usepackage{hyperref}
\usepackage{slashed}
\usepackage{array}

\hypersetup{colorlinks=true}
\hypersetup{linkcolor=violet}
\hypersetup{citecolor=violet}
\hypersetup{urlcolor=violet}

\numberwithin{equation}{section}

\usepackage
{datetime}
\usepackage{fancyhdr}
\usepackage{tikz, pgf}
\usetikzlibrary{shapes.misc}
\usetikzlibrary{shapes}
\usetikzlibrary{fit}
\usetikzlibrary{arrows}

\usetikzlibrary{decorations.pathmorphing}	
\usetikzlibrary{decorations.markings}
 \usetikzlibrary{patterns}

\tikzset{
    sugra/.style={decorate, decoration={snake}, draw=black},
    scalarphi/.style={dashed,draw=black, postaction={decorate},
        },
    scalarchi/.style={draw=brown}, 
    hwbou/.style={draw=blue, postaction={decorate}, ultra thick
        },
    vector/.style={draw=blue,decorate, decoration={snake}, draw},
	provector/.style={decorate, decoration={snake,amplitude=2.5pt}, draw},
	antivector/.style={decorate, decoration={snake,amplitude=-2.5pt}, draw},
   	 fermion/.style={draw=cyan, postaction={decorate},
        decoration={markings,mark=at position .55 with {\arrow[draw=black]{>}}}},
    fermionbar/.style={draw=cyan, postaction={decorate},
        decoration={markings,mark=at position .55 with {\arrow[draw=black]{<}}}},
    chspin/.style={draw=red, postaction={decorate},
        decoration={markings,mark=at position .55 with {\arrow[draw=black]{>}}}},
    chspinbar/.style={draw=red, postaction={decorate},
        decoration={markings,mark=at position .55 with {\arrow[draw=black]{<}}}},  
    fermionnoarrow/.style={draw=black},
    gluon/.style={decorate, draw=red,
        decoration={coil, amplitude=4pt, segment length=5pt}},
    scalar/.style={dashed,draw=black, postaction={decorate},
        decoration={markings,mark=at position .55 with {\arrow[draw=black]{>}}}},
    scalarbar/.style={dashed,draw=black, postaction={decorate},
        decoration={markings,mark=at position .55 with {\arrow[draw=black]{<}}}},
    electron/.style={draw=black, postaction={decorate},
        decoration={markings,mark=at position .55 with {\arrow[draw=black]{>}}}},
    scalarnoarrow/.style={dashed, draw=black},
    electron/.style={draw=black, postaction={decorate},
        decoration={markings, mark=at position .55 with {\arrow[draw=black]{>}}}},
	bigvector/.style={decorate, decoration={snake, amplitude=4pt}, draw},
    photon/.style={draw=violet, decorate, decoration={snake}, draw},
    higgs/.style={dashed, draw=black, postaction={decorate},
        },	
        goldstone/.style={draw=brown, postaction={decorate},
        },    
          ghost/.style={dashed, draw=blue, postaction={decorate},
        decoration={markings, mark=at position .55 with {\arrow[draw=black]{>}}}
        },  
          antighost/.style={dashed, draw=blue, postaction={decorate},
        decoration={markings, mark=at position .55 with {\arrow[draw=black]{<}}}
        }, 
            scalartwo/.style={dashed,draw=brown, postaction={decorate},
        decoration={markings,mark=at position .55 with {\arrow[draw=black]{>}}}},
    scalarbartwo/.style={dashed,draw=brown, postaction={decorate},
        decoration={markings,mark=at position .55 with {\arrow[draw=black]{<}}}}, 
    fermiontwo/.style={draw=purple, postaction={decorate},
        decoration={markings,mark=at position .55 with {\arrow[draw=black]{>}}}},
    fermionbartwo/.style={draw=purple, postaction={decorate},
        decoration={markings,mark=at position .55 with {\arrow[draw=black]{<}}}},    
        realscalar/.style={draw=black}, 
        fakerealscalar/.style={draw=white}, 
        realscalarone/.style={ draw=black},
    	realscalartwo/.style={draw=brown},    	    pseudoscalar/.style={draw=brown},
        mgluon/.style={decorate, draw=blue,
        	decoration={coil,amplitude=4pt, segment length=5pt}},
         weylfermion/.style={draw=orange, postaction={decorate},
        decoration={markings,mark=at position .55 with {\arrow[draw=black]{>}}}},
         weylfermionbar/.style={draw=orange, postaction={decorate},
        decoration={markings,mark=at position .55 with {\arrow[draw=black]{<}}}}, 
    majorana/.style={draw=cyan, postaction={decorate},
        decoration={markings,mark=at position .55 with {\arrow[draw=black]{><}}}},
    majoranabar/.style={draw=cyan, postaction={decorate},
        decoration={markings,mark=at position .55 with {\arrow[draw=black]{><}}}},    
   	wboson/.style={draw=blue,decorate, decoration={snake,amplitude=4pt}, draw},  
    zboson/.style={draw=violet, decorate, decoration={snake}, draw},   
    lepton/.style={draw=black, postaction={decorate},
        decoration={markings, mark=at position .55 with {\arrow[draw=black]{>}}}},
    leptonbar/.style={draw=black, postaction={decorate},
        decoration={markings, mark=at position .55 with {\arrow[draw=black]{<}}}}, 
    clepton/.style={draw=cyan, postaction={decorate},
        decoration={markings, mark=at position .55 with {\arrow[draw= black]{>}}}},
    cleptonbar/.style={draw=cyan, postaction={decorate},
        decoration={markings, mark=at position .55 with {\arrow[draw=black]{<}}}},   
   nlepton/.style={draw=orange, postaction={decorate},
        decoration={markings, mark=at position .55 with {\arrow[draw=black]{>}}}},
    nleptonbar/.style={draw=orange, postaction={decorate},
        decoration={markings, mark=at position .55 with {\arrow[draw=black]{<}}}},              
        graviton/.style={draw=blue, decorate, decoration={snake, amplitude=4pt}, draw},  
        spinj/.style={draw=red, decorate, decoration={snake, amplitude=4pt}, draw},  
        bgraviton/.style={draw=blue, decorate, decoration={snake, amplitude=4pt}, draw},  
        gravitino/.style={draw=red, postaction={decorate}, 
        decoration={snake,  markings, mark=at position .55 with {\arrow[draw=black]{><}}}},
    	gravitinobar/.style={draw=red, postaction={decorate},
        decoration={snake, markings, mark=at position .55 with {\arrow[draw=black]{><}}} },  
    phir/.style={draw=blue, postaction={decorate},},
   phil/.style={dashed,draw=blue,},
     phiav/.style={draw=cyan, postaction={decorate},},
   phidif/.style={dashed,draw=cyan,},  
    chir/.style={draw=red, postaction={decorate},},
   chil/.style={dashed,draw=red,},  
}

\newcommand{\treelevelthreepointskel}[4]{
\begin{scope}[shift={(0,0)}, rotate=#4]
	\draw[#1] (0,0)--(1.5,0);    
	\draw[#2][rotate=120] (0,0)--(1.5,0);  
	\draw[#3][rotate=-120](0,0)--(1.5,0);  

\end{scope}	 
}

\newcommand{\labeltreelevelthreepoint}[4]{
\begin{scope}[shift={(0,0)}, rotate=#4]

\node at (2,0) {#1}; 
\begin{scope}[shift={(0,0)},rotate=120]	
\node at (2,0) {#2};
\end{scope}	
\begin{scope}[shift={(0,0)},rotate=240]	
	\node at (2,0) {#3};
\end{scope}	
	
\end{scope}	 
} 

\newcommand{\treefourpointexchange}[6]{
\begin{scope}[rotate=#6]

\begin{scope}[shift={(-.75,0)} ]

\treelevelthreepointskel{fakerealscalar}{#1}{#2}{0}	
\begin{scope}[shift={(1.5,0)}]	
\treelevelthreepointskel{#5}{#3}{#4}{180}	
\end{scope}	

\end{scope}	 
\end{scope}	 
}

\newcommand{\treefourpointcrossexchange}[6]{
\begin{scope}[rotate=#6]

\begin{scope}[shift={(-.75,0)} ]

	\draw[#5] (0,0)--(1.5,0);    
	\draw[#2][rotate=-120](0,0)--(1.5,0); 

	\draw[#4][rotate=30](0,0)--(2.6,0); 
	\filldraw[white] (.75,.45) circle (.2);  

\begin{scope}[shift={(1.5,0)}, rotate=0]
	\draw[#3][rotate=-60] (0,0)--(1.5,0);
	
	\draw[#1][rotate=150](0,0)--(2.6,0);        
\end{scope}	 

\end{scope}	 
\end{scope}	 
}

\newcommand{\labeltreelevelfourpoint}[5]{
\begin{scope}[shift={(0,0)}, rotate=#5]

\node at (2,0) {#1}; 
\begin{scope}[shift={(0,0)},rotate=90]	
\node at (2,0) {#2};
\end{scope}	
\begin{scope}[shift={(0,0)},rotate=180]	
	\node at (2,0) {#3};
\end{scope}	

\begin{scope}[shift={(0,0)},rotate=270]	
	\node at (2,0) {#4};
\end{scope}	
	 
\end{scope}	 
}

\newcommand{\labeltreefourpointexchange}[6]{
\begin{scope}[rotate=#6]

\begin{scope}[shift={(-.75,0)} ]

\node at (.75,-.5) {#5}; 

\labeltreelevelthreepoint{}{#1}{#2}{0} 
\begin{scope}[shift={(1.5,0)}]	
\labeltreelevelthreepoint{}{#3}{#4}{180}
\end{scope} 
	
\end{scope}	 
\end{scope}	 
}

\newcommand{\treelevelfourpointskel}[5]{
\begin{scope}[shift={(0,0)}, rotate=#5]
	\draw[#1] (0,0)--(1.5,0);     
	\draw[#2][rotate=90] (0,0)--(1.5,0);    
	\draw[#3][rotate=180] (0,0)--(1.5,0);   
	\draw[#4][rotate=270] (0,0)--(1.5,0);

\end{scope}	 
}

\usepackage{comment}
\definecolor{cobalt}{rgb}{0.0, 0.28, 0.67}

\specialcomment{rudranote}{%
  \begingroup \color{cobalt}}{%
  \endgroup}

\tikzstyle{block} = [draw, rectangle, 
    minimum height=3em, minimum width=6em]

\def\gem{ \textrm{g}_{\textrm{em}} }

\newcommand{\gcou}{{\textrm{g}}}

 \definecolor{ao(english)}{rgb}{0.0, 0.5, 0.0}
\usepackage{pifont} 
\newcommand{\xmark}{\ding{55}} 
\newcommand{\cmark}{\ding{51}}

\usepackage{comment}

 \specialcomment{rudradetails}{%
  \begingroup \color{blue}}{%
  \endgroup} 

\newcommand{\coue}{\textrm{e}}

\newcommand{\spin}{\mathcal{S}}

\newcommand{\spp}[2]{(#1 \cdot #2)}

\def\iimg{ {\bf i}}

\def\diag{\textrm{diag}}

\def\invc{\textrm{in}}
\def\outvc{\textrm{out}}

\hbadness=\maxdimen
\vbadness=\maxdimen

\allowdisplaybreaks[1]

\title{\boldmath Compton amplitude for massive bosons of arbitrary spin}

\vspace{10mm}

\author[\symbishop]{Aakash Kumar}
\author[\symbishop]{, Arnab Rudra}
\author[\symbishop,\symknight]{, Manav Shah }
\author[\symbishop]{, Rahul Shaw}
\author[]{\\ }
\vspace*{4.0ex} 
\vspace{10mm}
\affiliation[\symbishop]{Indian Institute of Science Education and Research Bhopal,\\
 Bhopal Bypass Road, Bhauri, Madhya Pradesh 462066, India.\\ }
\vspace*{4.0ex} 
\affiliation[\symknight]{Department of Applied Mathematics and Theoretical Physics,\\
Wilberforce Road, Cambridge CB3 0WA, UK.\\ }
\vspace*{4.0ex} 
\vspace*{2.0ex}
\emailAdd{aakash19@iiserb.ac.in}
\emailAdd{rudra@iiserb.ac.in} 
\emailAdd{mns48@cam.ac.uk}
\emailAdd{rahulshaw.phy@gmail.com}  

\abstract{In this work, we write down an analytic expression of electromagnetic tree-level Compton amplitude for a completely symmetric traceless (bosonic) higher spin particle in any dimension. Our analysis is restricted to the three point function which is unique in the Infrared and responsible for the Coulomb interactions/soft photon theorem. We propose an analogue of $R_\xi$ gauge in a theory of higher spin particles. We demonstrate that the theory is unitary only for $\xi=1,\infty$.}

\begin{document} 
\maketitle

\newpage
\section{Introduction}
\label{sec:krsscomptonIintro}
In 1923, Arthur Holly Compton observed that a high-frequency light/high energy photon $\gamma$ can scatter from an electron $e^-$: 
\begin{equation}
	e^-+\gamma\longrightarrow e^-+\gamma
\end{equation}
This experiment demonstrate particle-nature of light. This was a very important discovery\footnote{Arthur Holly Compton was awarded Nobel prize (along with C. T. R. Wilson) in 1927 for this discovery.}. This scattering process is known as the Compton scattering. Later, this result was reproduced using QED Feynman diagrams and it provided an important check for QED. 
\begin{figure}[h]
\begin{center}
\begin{tikzpicture}[line width=1.5 pt, scale=1.3]
 
\begin{scope}[shift={(0,0)}]	
		
\treefourpointexchange{fermion}{photon}{photon}{fermionbar}{fermion}{0}	

\labeltreefourpointexchange{ $k_2$ }{ $k_3$ }{ $k_4$ }{ $k_1$ }{}{0} 

\draw (0,-2.5) node {$t-$channel};
\end{scope}

\begin{scope}[shift={(7,0)}]	
		
\treefourpointcrossexchange{photon}{fermionbar}{fermion}{photon}{fermion}{180}	

\labeltreefourpointexchange{ $k_2$ }{ $k_3$ }{ $k_4$ }{ $k_1$ }{}{0} 
\draw (0,-2.5) node {$u-$channel};

\end{scope}

\end{tikzpicture} 
\end{center}
\caption{Tree-level Feynman diagrams for compton scattering in Spinor QED}
\label{fig:comptonspinorqed}
\end{figure}
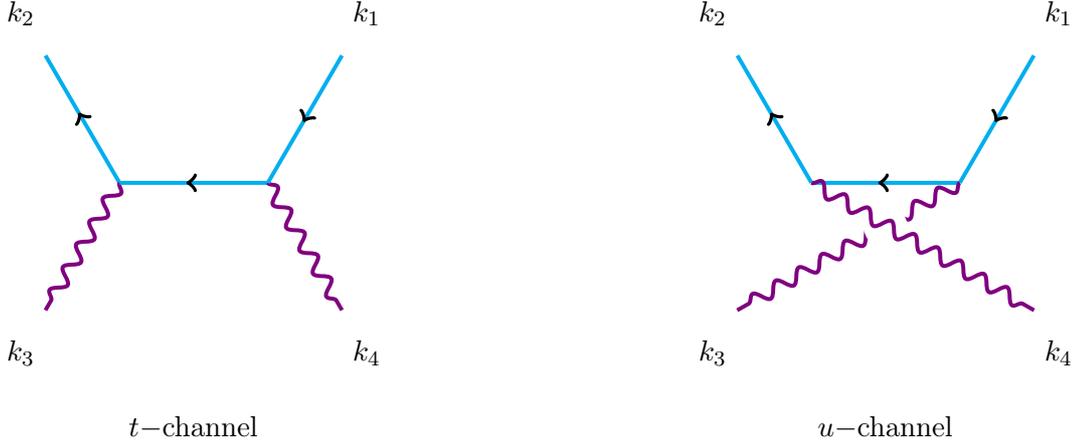

In this note, we refer to a process of the following form as (generalized) Compton amplitude
\begin{equation}
	\Phi+\gamma\longrightarrow \Phi+\gamma 
\end{equation}
Here $\Phi$ is a charged massive particle of mass $m$ with integer spin (i.e. a Bosonic higher spin particle). The goal of this note is to compute the Compton amplitude for spinning massive particles, where the spin can take any integer value. 

\paragraph{Restatement of the problem (using QFT languages)} Consider a theory of a massive particle interacting with photons in 3+1 dimensional Minkowski space. In 1936, Paul Dirac attempted to write down Wave equation(/Dirac equation) for spins greater than a half \cite{Dirac:1936tg}.  This was followed up by Fierz and Pauli in precense of electromagnetic field \cite{Fierz:1939ix} and later followed up by many people \cite{Singh:1974qz, Singh:1974rc}. The problem turns out to be notoriously difficult in the Lagrangian approach. We approach this problem using scattering amplitude and the on-shell method. The on-shell method is a popular technique in the recent hep-th literature. In our context, we mean the following: We do not write down a Lagrangian. We directly write down the 3-point Feynman rules, which satisfy all the necessary properties (like reality, decoupling of pure gauge polarisation) when external legs are on-shell. With those Feynman rules, we construct the higher point functions following QFT rules and again check whether the amplitude satisfies all the necessary properties (cutting rules, decoupling of unphysical d.o.f) or not. If any property is lacking, we add appropriate contact interaction to reestablish the property. Sometimes, it is not possible to restore some of the properties. Then, we declare it to be a bad theory; otherwise, we consider the theory and subject it to consistencies at a higher-point amplitudes and higher loop corrections. In 3+1 dimensions, spinor helicity variables are mostly used to implement the on-shell method \cite{Elvang2013, ArkaniHamed:2012nw}, even though the idea is also implementable in terms of usual QFT variables: Lorentz vectors and tensors. In this work, we will be following the second method. 

We are interested in finding a model of massive higher spin\footnote{Any massive particle with spin greater one will be denoted as a massive higher spin particle in this work} particle(s) weakly interacting with the photons and we will be using on-shell approach for this problem.  

We denote the polarisation of a photon with momenta $k_\mu$ as $\epsilon_\mu (k_\mu) $; it is transverse to the momenta ($k\cdot \epsilon(k)=0$). The other key character of this play is a massive spin-$\spin$ particle. For the moment, we restrict to only bosonic particles. In $D$ spacetime dimensions, the massive bosonic particles are representations of the little group $SO(D-1)$; thus, the bosonic particles are in one-to-one correspondence with Young-Tableaux of $SO(D-1)$. We make another simplifying assumption. We restrict to completely symmetric Young Tableaux only. This means that the field corresponding to the higher spin particle is symmetric, traceless and (on-shell)transverse. The fields in momentum space satisfy the following conditions 
\begin{equation}
	\Phi_{\mu_1\cdots\mu_\spin}
\qquad:\qquad 	
\Phi_{\mu_1\cdots\mu_\spin}=\Phi_{(\mu_1\cdots\mu_\spin)}
\quad,\quad 
\eta^{\mu_1\mu_2}
\Phi_{\mu_1\cdots\mu_\spin}=0
\quad,\quad 
k^{\mu_1}
\Phi_{\mu_1\cdots\mu_\spin}=0
\label{comptonproblemstatement1}
\end{equation} 
 We are interested in writing down the four-point amplitude of two $\Phi$s and two photons. We adopt the convention that the two massive particles are assigned labels 1 and 2, and the two photons are assigned labels $3$ and $4$. We denote Compton amplitude as 
\begin{equation}
	\mathcal{M}(\Phi_1,\bar\Phi_2,\varepsilon_3,\varepsilon_4 )
\label{comptonproblemstatement2}
\end{equation}
We want to find $\mathcal{M}(\Phi_1,\bar\Phi_2,\varepsilon_3,\varepsilon_4 )$ such that 
\begin{enumerate}
	\item Non-analytic behaviour of the amplitude is, at most, a pole (i.e. in the language of perturbative QFT, we restrict our attention to tree-level amplitudes). 
	\item The amplitude is gauge invariant 
\begin{equation}
\mathcal{M}(\Phi_1,\bar\Phi_2,\varepsilon_3=k_3,\varepsilon_4 )=0=\mathcal{M}(\Phi_1,\bar\Phi_2,\varepsilon_3,\varepsilon_4=k_4 )	
\label{comptonproblemstatement3}
\end{equation} 
	
	\item The amplitude satisfies the Cutkosky cutting rules (i.e. in the language of perturbative QFT, the amplitude is unitary).  
\end{enumerate}
We want to address this problem for any integer spin in any spacetime dimensions.  

\paragraph{Review of recent literature} In recent literature, there has been a lot of work on Compton amplitude due to its potential application to gravitational waves. Since the discovery of gravitational wave \cite{LIGOScientific:2016aoc,LIGOScientific:2017vwq,LIGOScientific:2018mvr}, there has been a lot of effort in understanding the gravitational waveform and radiation loss in binary systems (bound states and also scattering). Essentially, the interest is in understanding the two-body Hamiltonian in general relativity. Feynman amplitudes have turned out to be very useful in understanding the two-body Hamiltonian \cite{Damour:2017zjx, Bjerrum-Bohr:2018xdl}. Consider two massive bodies of masses $m_1$, $m_2$ and radii $r_1$ and $r_2$; if any of the bodies is a black hole, then $r$ is the Schwarzschild radius. We are considering their gravitational dynamics at separation $R$. The fundamental constants in the problem are $G$, $c$ and $\hbar$. The quantum effects are controlled by Compton Wavelength
\begin{equation}
	\lambda_i=\frac{\hbar}{m_i c}
\label{pparamater1}	
\end{equation}
as long as the separation $R$ is much much greater than $\lambda_i$, we can treat the $i$th particle classically
\begin{equation}
	R/\lambda_i\gg 1 \implies \textrm{No quantum Effect}
\label{pparamater3}	
\end{equation}
In our case, we will be in the classical regime. 

Then we consider the two more dimensionless numbers (which is essentially the radius of the objects divided by the separation between the two objects)
\begin{equation}
	\frac{r_1}{R}\qquad,\qquad \frac{r_2}{R}
\label{pparamater5}	
\end{equation}
If both of these numbers are very small, then we can treat the bodies as point-like bodies with multipoles. The Third dimensionless quantity is the Schwarzschild radius divided by the separation 
\begin{equation}
	\lambda_i^{(S)}=\frac{Gm_i}{c^2 R}
\label{pparamater7}	
\end{equation}
As long as the separation is much greater than the Schwarzschild radii, we can ignore or treat the general relativistic effects perturbatively.  
 
The other dimensionless quantity that we can construct is 
\begin{equation}
	\frac{v_i^2}{c^2}
\label{pparamater9}	
\end{equation}
This dimensionless number controls the special relativistic correction. If this number is very small, then it is the Galilean/non-relativistic limit. 

When both the parameters(\eqref{pparamater7} and \eqref{pparamater9}) are small, and the observable quantities for a bound state are written as expansion in  \eqref{pparamater9}, it is called the post Newtonian(PN) expansion. Instead when \eqref{pparamater9} is large but \eqref{pparamater7} is small and the observable quantities are written as expansion in \eqref{pparamater7}, it is called the post Minkowskian(PM) expansion. In the language of QFT, we consider a massive quantum particle coupled to the graviton. Then, we take the classical limit and compute the potential from the Feynman diagrams. From the Compton amplitude, we can construct the
Feynman diagram scattering of four massive particles due to two graviton exchanges. This would help us find the gravitational potential for higher PM order. In the first approximation, one can ignore the spin\footnote{We are studying extended bodies. Here, spin can be understood as the angular momentum around its own axis.}, i.e. study the two-body problem of two spin-less bodies in GR. This problem can mapped to the problem of Scalar fields coupled to Einstein's GR\cite{Bjerrum-Bohr:2013bxa,Neill:2013wsa}. 

One natural question is to understand the dynamics of Kerr Black holes \cite{Vaidya:2014kza}. This problem can be mapped to the scattering of higher spin particles. Let $\spin$ be the spin quantum number. One is essentially interested in $\spin\rightarrow\infty$ limit keeping $\spin\hbar$ fixed and identify it as the classical spin \cite{Guevara:2017csg, Cangemi:2022abk, Cangemi:2024apk, Pichini:2023cqn}.

Due to this correspondence, there has been a lot of effort in understanding higher spin Compton amplitude \cite{Vines:2017hyw, Bini:2018ywr, Guevara:2018wpp, Chung:2018kqs, Guevara:2019fsj, Chiodaroli:2021eug, Bautista:2021wfy, Cangemi:2022bew, Cangemi:2023ysz, Cangemi:2023bpe, Bautista:2022wjf, Bautista:2024agp}\footnote{The literature on higher spin compton and black hole physics is huge. In case we have missed some important references, we request the reader to point it out to us.} fueled by the discovery of Spinor-Helicity formalism \cite{Arkani-Hamed:2017jhn, Arkani-Hamed:2019ymq} and extracting the classical physics \cite{Kosower:2018adc} out of it. In fact, one can also ask the same question in the context of electromagnetism \cite{Bern:2023ccb, Bini:2024pdp, Akhtar:2024mbg} and try to understand it as a first step towards understanding the gravitational physics problem. 

\subsection{Main results}
\label{sec:krsscomptonIresults}

The analysis in this work is computationally long and involved. Here, we summarise the main results of the paper for better readability:
\begin{enumerate}
	\item We would be considering higher spin particles in Minkowski spacetime with dimension $D$ ($D>3$). In this paper, we restrict to higher spin particles, which are symmetric, traceless Lorentz tensors (given in \eqref{comptonproblemstatement1}). The on-shell condition on these particles (in the momentum space) is given by 
\begin{equation}
	(k^2+m_{\spin}^2)\Phi_{\mu_1\cdots \mu_\spin}=0
\qquad,\qquad 
k^{\mu_1}\Phi_{\mu_1\cdots \mu_\spin}=0
\label{krsscomptonIresults1}
\end{equation}
\item For spin-$\spin$ greater than 1, we do not have any Lagrangian which imposes all the above conditions. We write down the propagator from the on-shell considerations \cite{Ingraham:1974un}
\begin{equation}
\langle\Phi_{\mu_1\cdots \mu_\spin}(p)\bar{\Phi}_{\nu_1\cdots \nu_\spin}(-p)\rangle=\frac{\iimg}{p^2+m^2-\iimg \varepsilon}	\mathcal{P}^{(\spin)}_{\mu_1\cdots \mu_\spin,\nu_1\cdots \nu_\spin}(p)
\label{krsscomptonIresults3}
\end{equation}
$\mathcal{P}^{(\spin)}_{\mu_1\cdots \mu_\spin,\nu_1\cdots \nu_\spin}(p)$ is known as the projector; it is symmetric and traceless in the $\mu $ indices and in the $\nu$ indices. It can be written in terms of the rank-2 tensor $\Theta_{\mu\nu}$, which is defined as
\begin{equation}
	\Theta_{\mu\nu}(p)=\eta_{\mu\nu}-\frac{p_\mu p_\nu}{p^2}
\label{krsscomptonIresults5}
\end{equation}
the explicit expression of $\mathcal{P}^{(\spin)}_{\mu_1\cdots \mu_\spin,\nu_1\cdots \nu_\spin}(p)$ can be found in \eqref{raamhspinreview4.1}.

\item After writing down the propagator, we find interaction terms. We write down on-shell the point vertex of two higher spin particles and a massless spin particle (/photon). We restrict to the three-point function, which gives rise to Coulomb interaction ($1/r^{D-3}$ potential\footnote{We are assuming $D>3$.}) in the long distance. Thus, this interaction is the dominant interaction at low energy/long distance (i.e in the infrared). The three-point function is given by (we denote the massive particles with indices 1 \& 2 and the massless particles with index 3)
\begin{equation}
	\iimg \frac{1}{2m}(\Phi^{\mu_1\cdots \mu_\spin}(k_1)\bar\Phi_{\mu_1\cdots \mu_\spin}(k_2))\varepsilon_3\cdot k_{12}
\qquad,\qquad
k_{12}=k_1-k_2	
\label{krsscomptonIresults7}
\end{equation}
$m$ is the mass of the massive particles.

\item We compute tree-level Compton amplitude using the propagator given in \eqref{krsscomptonIresults3} and the three-point function given in \eqref{krsscomptonIresults7}. In our convention (as depicted in \ref{fig:comptonspinorqed}), massive states are labelled with index 1\& 2 and the massless states are labelled with indices 3\& 4. The tree-level compton amplitude $	\widetilde{\mathcal{M}}$ gets contributions from two channels: $t$-channel contribution ($\widetilde{\mathcal{M}}^{(t)}$) and $u$-channel contribution ($\widetilde{\mathcal{M}}^{(u)}$). 
\begin{equation}
		\widetilde{\mathcal{M}}=	\widetilde{\mathcal{M}}^{(t)}+	\widetilde{\mathcal{M}}^{(u)}
\end{equation}

\item 

We found that the contributions of these channels do not vanish if we put pure gauge polarisation for the external photons. 
\begin{eqnarray}
		\widetilde{\mathcal{M}}\Big|_{\varepsilon_3\rightarrow k_3}\ne 0
\qquad,\qquad	 
		\widetilde{\mathcal{M}}\Big|_{\varepsilon_4\rightarrow k_4}\ne 0
\end{eqnarray}
Moreover, the non-invariant terms also have massless poles. Thus the total contributions from the exchange diagrams is not gauge-invariant.  

\item 
For spin-$1$ massive bosons, it is possible to get rid of these terms by adding an extra massless scalar. This is not surprising as these scalars appear in the Higgs mechanism for massive spin-$1$. The situation is similar to the $\xi=0$ for $R_\xi$ gauge where there are massless would-be-goldstone bosons. These would-be-goldstone bosons are also important to ensure unitarity.

\item Motivated from the spin-$1$ example, we attempted to cure the problem by adding massless particle(s). We observed that for spin greater than one, it is not possible for extra massless particle(s) to restore unitarity. 

\item Subsequently (again motivated from the spin-$1$ example), we propose\footnote{The proposal is ad-hoc; we do not provide any justification in terms of Lagrangian or some other principle.} the analogue of $R_\xi $ gauge for massive higher spin particles. We basically then replace the expression of $\Theta$ (in \eqref{krsscomptonIresults5}) by the following expressions
\begin{equation}
	\Theta_{\mu \nu}(k,\xi)=\eta_{\mu \nu} +\frac{k_\mu k_\nu(\xi-1) }{k^2+\xi m^2-\iimg \epsilon}
\qquad,\qquad \xi \in [0,\infty)	
\end{equation} 
This modifies the expression of the propagator. We compute the Compton amplitude again with this new propagator. We find the amplitude is still not gauge-invariant. However, it is possible to restore gauge invariance only for two values of $\xi$ 
\begin{itemize}
	\item $\xi=\infty$: In this case, the non-invariant part of the amplitude is analytic. We showed that the non-invariance can be removed by local contact term $\widetilde{\mathcal{M}}^{(\textrm{ct})}$. We define the full amplitude to be 
\begin{equation}
	\widehat{\mathcal{M}}=\widetilde{\mathcal{M}}- \widetilde{\mathcal{M}}^{(\textrm{ct})}
\end{equation} 
$\widehat{\mathcal{M}}$ is gauge invariant  
\begin{equation}
	\widehat{\mathcal{M}}\Big|_{\varepsilon_3\rightarrow k_3}
	=0=\widehat{\mathcal{M}}\Big|_{\varepsilon_4\rightarrow k_4}
\label{spinzerocompton7}	
\end{equation}
Note that the contact is not unique. The role of this contact term is to cancel the gauge-invariance of the exchange contributions. Thus given a contact term we can always construct another contact term which differs from the first one by a local gauge invariant piece. This is the only ambiguity in constructing the contact term. However, the gauge invariance is constructed out of linearized field strength and thus gauge invariant contact term often has more derivatives than the gauge non-invariant term. For example consider the scalar QED. In this case, the contact term is 
\begin{eqnarray}
	\iimg (\varepsilon_3\cdot \varepsilon_4)
\label{ctambiguity1}	
\end{eqnarray}
We can always a term of the following form to the above term
\begin{eqnarray}
	\iimg f(s,t,u)\,  \mathcal{B}_3\cdot \mathcal{B}_4
\label{ctambiguity2}	
\end{eqnarray}
 where $\mathcal{B}_{\mu\nu}^{(i)}=k_\mu^{(i)} \epsilon_\nu^{(i)}-k_\nu^{(i)} \epsilon_\mu^{(i)}$. We can see that it has at least 2 more derivative than \eqref{ctambiguity1}.

	\item $\xi=1$: In this case, again, the gauge non-invariant part is analytic and can be removed by contact term. However, the amplitude obtained after subtracting the contact term does not obey the cutting rules/perturbative unitarity. In order to restore that, we need to add particles of mass $m$ (i.e., the same as the mass of the massive higher particles) and spin ranging from 0 to $\spin-1$. For spin-$1$ particles, these are the would-be-goldstone bosons. We denote all such unphysical particles by $\varphi$; we also determine the precise 3-point vertex  that is needed to ensure unitarity. 

Note that the string theory answers are available for $\xi=1$ (Seigel gauge), and the feature we see is similar to string theory. 
\end{itemize}
For any other value of $\xi$, it is not possible to restore gauge invariance/unitarity. 
\end{enumerate}

\paragraph{Organization of the paper } We start by reviewing the basics of bosonic higher spin particles in sec \ref{subsec:krsscomptonIbhsreview}. Then, we briefly discussion about three-point functions in sec \ref{subsec:krsscomptonIthreepoint}. In section \ref{sec:krsscomptonIscalar}, we compute Compton scattering for spin-$0$ and spin-$1$ particles. The results are known using the Lagrangian approach. This section serves the purpose of explaining the basic logic that is implemented in this paper to understand Compton amplitude for higher spin particles; we also discuss the extension of $R_\xi$ gauge to higher spin particles in this section. In section \ref{sec:krsscomptonIspintwo}, we implement the logic to the spin-$2$ particles for $\xi=\infty$ and show how to find gauge invariant Compton amplitude. Currently, it is not known how to reproduce this result using the Lagrangian approach. In section \ref{sec:krsscomptonIarbitraryspin}, we extend the result arbitrary spins and write down a manifestly gauge invariant answer for the Compton amplitude for $\xi=\infty$. We check the Cutkosky cutting rule in sec \ref{sec:krsscomptonIunitarity}. While $\xi=\infty$ satisfies cutting rule, $\xi=1$ does not. In the latter case, unitarity is restored after adding extra particles with very particular three-point vertices. In section \ref{sec:krsscomptonIconclusion}, we discuss the applicability of the method to more general problems. We also list a few open problems that we want to attempt in the near future.

\section{Basics}
\label{sec:krsscomptonIreview}
In this section, we review the material that exists in the literature and is useful to us throughout the paper. 

\subsection{Bosonic higher spin}
\label{subsec:krsscomptonIbhsreview}

We begin by summarizing higher spin particles. In the literature, any particle with a spin of more than two is known as a Higher spin particle. In an interacting theory, it is possible to show that higher spin particles cannot be massless. A massive particle in the Minkowski space is characterized by the representation of the little group. For our purposes, we restrict to the completely symmetric and traceless representation of the little group. We have already summarized the properties of this kind of particle in \eqref{comptonproblemstatement1}. 
The equation motion for the massive particle is 
\begin{equation}
    (\Box +m_\spin^2)\Phi_{\mu_1\cdots\mu_\spin} = 0
\label{raamhspinreview1}
\end{equation}
Now we write down the propagator for higher spin particle
 \cite{Ingraham:1974un, Balasubramanian:2021act}. The expression of it involves a projector (see eqn \eqref{krsscomptonIresults3}) which is given by 
\begin{eqnarray}
    \mathcal{P}^{(\spin)}_{(\mu),(\nu)}(p)=
    \Bigg[\sum_{a=0}^{\lfloor \spin/2\rfloor}
    A(\spin,a,D)\, 
    \Theta_{\mu_1\mu_2}(p)\Theta_{\nu_1\nu_2}(p)\cdots\Theta_{\mu_{2a-1}\mu_{2a}}(p)\Theta_{\nu_{2a-1}\nu_{2a}}(p)\Theta_{\mu_{2a+1}\nu_{2a+1}}(p)
    \dots\Theta_{\mu_\spin\nu_\spin}(p)\Bigg]_{\text{sym} (\mu,\nu)} 
\label{raamhspinreview4.1}
    \end{eqnarray}    
    where 
    \begin{subequations}
    \begin{eqnarray}
        \Theta_{\mu\nu}(p)&=&\eta_{\mu\nu}-\frac{p_\mu p_\nu}{p^2}
\label{raamhspinreview5a}
    \\    
        \lfloor \spin/2\rfloor &=& 
        \textrm{greatest integer smaller than } \spin/2
\label{raamhspinreview5b}
    \end{eqnarray}    
    \end{subequations}
    $A(\spin,a,D)$ is given by 
    \begin{equation}
        A(\spin,a,D)=\frac{(-1)^a \spin!(2\spin+D-2a-5)!!}{2^a a!(\spin-2a)!(2\spin+D-5)!!}
\label{raamhspinreview11}
    \end{equation}
    $D$ is the number of spacetime dimensions. One can check that the projector satisfies the following properties
\begin{itemize}
	\item Transverse to the momentum
\begin{equation}
	p^{\mu_1}\mathcal{P}^{(\spin)}_{\mu_1\cdots \mu_\spin;\nu_1\cdots \nu_\spin}(p)=0=p^{\nu_1}\mathcal{P}^{(\spin)}_{\mu_1\cdots \mu_\spin;\nu_1\cdots \nu_\spin}(p)
\label{raamhspinreview12}
\end{equation}

	\item Traceless in the $\mu$ indices and in the $\nu $ indices
\begin{equation}
	\eta^{\mu_1\mu_2}\mathcal{P}^{(\spin)}_{\mu_1\mu_2\mu_3\cdots \mu_\spin;\nu_1\cdots \nu_\spin}(p)=0=\eta^{\nu_1\nu_2}\mathcal{P}^{(\spin)}_{\mu_1\cdots \mu_\spin;\nu_1\nu_2\nu_3\cdots \nu_\spin}(p)
\label{raamhspinreview13}
\end{equation}
	\item It is an idempotent operator 
\begin{equation}
	{\mathcal{P}^{(\spin)}_{\mu_1\cdots \mu_\spin}}^{\nu_1\cdots \nu_\spin}{\mathcal{P}^{(\spin)}_{\nu_1\cdots \nu_\spin}}^{\rho_1\cdots \rho_\spin}={\mathcal{P}^{(\spin)}_{\mu_1\cdots \mu_\spin}}^{\rho_1\cdots \rho_\spin}
\label{raamhspinreview14}
\end{equation}
	\item Trace in one $\mu$ and one $\nu $ index satisfies the following property(proved in appendix \ref{traceproperty})
\begin{equation}
	\eta^{\mu_\spin\nu_\spin}\mathcal{P}^{(\spin)}_{\mu_1\cdots \mu_\spin;\nu_1\cdots \nu_\spin}(p)=\frac{\textrm{dof}(\spin,D)}{\textrm{dof}(\spin-1,D)}\mathcal{P}^{(\spin-1)}_{\mu_1\cdots \mu_{\spin-1};\nu_1\cdots \nu_{\spin-1}}(p)
\label{raamhspinreview15}
\end{equation}
where $\textrm{dof}(\spin, D)$ is the number of independent degrees of freedom of the Quantum particle representation the quantum field which is a symmetric traceless transverse tensor of rank $\spin$ in $D$ spacetime dimensions
\begin{equation}
	\textrm{dof}(\spin,D)= {{D+\spin-2}\choose{\spin}}- {{D+\spin-4}\choose{\spin-2}}
\label{raamhspinreview21}
\end{equation}
\end{itemize}
The diagramming notation for the higher spin propagator is denoted in fig. \ref{fig:krssIfeynmanrule1}
\begin{figure}[h]
 
\begin{center}
	\begin{tikzpicture}[line width=1.5 pt, scale=2] 


	\begin{scope}[shift={(0,0)}]	
	\draw[chspin] (0,0)--(1,0);
	 \node at (1.5,0) {$\Phi_{(\nu)} $};	
	  \node at (-.5,0) {$\Phi_{(\mu)} $};		
	  \node at (.5,-.25) {$ p \rightarrow $};		
  \node at (3.5,0) {$\frac{-\iimg}{p^2+m^2-\iimg\varepsilon} \mathcal{P}^{(\spin)}_{(\mu),(\nu)}(p)  	$};	     		
	\end{scope}	 
  
	 \end{tikzpicture}	
\end{center}
\caption{Propagators for the MHS}
\label{fig:krssIfeynmanrule1} 
\end{figure}

\paragraph{A handy-way to write higher spin polarisation compactly:} Now we introduce a short-hand notation to write the polarizations of massive spin-$\spin$ particles. For our purposes, it is useful to define auxiliary variables $\zeta_\mu$s such that we write
\begin{equation}
	\Phi_{\mu_1\cdots \mu_\spin}\simeq 	\zeta_{\mu_1}\cdots \zeta_{\mu_\spin}
\label{raamhspinreview2}
\end{equation}
Then the traceless property of $\Phi$ implies that $\zeta$ is a null vector
\begin{equation}
\eta^{\mu_1\mu_2}
\Phi_{\mu_1\cdots\mu_\spin}=0
\implies 
\zeta\cdot \zeta= 0	
\label{raamhspinreview3}
\end{equation}
and the transverse nature of $\Phi$ implies transverse nature of $\zeta$
\begin{equation}
	k^{\mu_1}
\Phi_{\mu_1\cdots\mu_\spin}=0
\implies 
k^{\mu}\zeta_{\mu}=0
\label{raamhspinreview4}
\end{equation}
The sole reason to introduce \eqref{raamhspinreview2} is to write big expressions in a compact way. For example, consider the following expression
\begin{eqnarray}
\Phi_{\mu_1\cdots \mu_\spin} k^{\mu_1}\cdots k^{\mu_\spin}
	\simeq (\zeta\cdot k)^{\spin}
\end{eqnarray}
This is a very useful tool to compacttly write the expressions; we can always go back to the original expression by applying the following derivative operator
\begin{equation}
	\frac{1}{\spin !}\Phi_{\mu_1\cdots\mu_\spin}\frac{\partial^\spin}{\partial\zeta_{\mu_1}\cdots\partial\zeta_{\mu_\spin}}
\end{equation}

\subsection{Three point functions}
\label{subsec:krsscomptonIthreepoint}

Now, we discuss the three-point function of two massive spin-$\spin$ particles and one photon. We are choosing three points which are on-shell gauge-invariant (and not necessarily gauge-invariant off-shell). Moreover, if two three-point functions agree on-shell, we declare them to be the same. There are $2\spin+1$ independent structures \cite{Arkani-Hamed:2017jhn}. We will not go into details about that since, in this work, we focus on the three-point function, which is completely unique in the IR 
\begin{equation}
		-\iimg\frac{\gem}{2m} (\zeta_1\cdot\bar \zeta_2)^\spin(\varepsilon_3\cdot k_{12})
\label{raamhspinthreepointfunction1}
\end{equation}
where $k_{12}^\mu=k_1^\mu-k_2^\mu$. This is the unique three-point function in the IR, which leads to Coulombic forces in the long distance. The soft photon theorem is determined by this three-point function \cite{Weinberg1973}. The other three-point functions take the following form 
\begin{eqnarray}
V(\spin,n)&=&	-\iimg \frac{\coue_{2n+1}}{2m^{2n+1}}(\bar \zeta_1\cdot\zeta_2)^{\spin-n}(\bar \zeta_1\cdot k_{3})^n( \zeta_2\cdot k_{3})^n(\varepsilon_3\cdot k_{12})
\label{raamhspinthreepointfunction2}
\\
U(\spin,n)&=&		-\iimg\frac{\coue_{2n}}{m^{2n+1}}(\bar \zeta_1\cdot \zeta_2)^{\spin-n}(\bar \zeta_1\cdot k_{3})^{n-1}(\zeta_2\cdot k_{3})^{n-1}\Big( (\zeta_2\cdot \varepsilon_3)(\bar \zeta_1\cdot k_{3})-(\bar \zeta_1\cdot \varepsilon_3)(  \zeta_2 \cdot k_{3})\Big)\qquad
\end{eqnarray}
where $n=1,\dots, \spin$ and $e_1$ is reserved for $\gem$. So, the total most general three-point function is given by
\begin{equation}
	\mathcal{V}^{(\spin)}=\sum_{n=0}^\spin V(\spin,n)+\sum_{n=1}^{\spin} U(\spin,n)
\label{raamhspinthreepointfunction4}
\end{equation}
We briefly discuss the discrete symmetries ($C$, $R$ \& $T$) of the theory. The interactions are manifestly invariant under time reversal and reflection. The charge-conjugation follows from the following the action of the charge conjugation transformation on the higher spin particle \cite{Chakraborty:2024wdg}
\begin{equation}
C\qquad:\qquad 		\Phi_{\mu_1\cdots\mu_\spin}\longrightarrow 	\bar{\Phi}_{\mu_1\cdots\mu_\spin}
\label{raamhspinthreepointfunction5}
\end{equation}  
In 2+1 dimensions (and in other odd dimensions), it is possible to write down three-point functions \cite{Kumar:2024Aug} which violates these discrete symmetries. However, we do not consider them in this work. In this work, we restrict ourselves to \eqref{raamhspinthreepointfunction1}

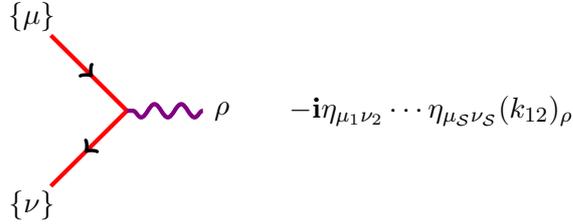
\begin{figure}[h]
\begin{center}
\begin{tikzpicture}[line width=1 pt, scale=1]

\begin{scope}[shift={(0,0)}, rotate=0]


\draw [ultra thick, chspin] (-1,1) -- (0,0);
\draw [ultra thick, chspinbar] (-1,-1) -- (0,0); 
\draw [ultra thick, photon] (1,0) -- (0,0);

\draw (-1.25,1.25) node {$\{\mu\}$};
\draw (-1.25,-1.25) node {$\{\nu\}$};
\draw (1.25,0) node {$\rho$};

  \node at (4,0) {$-\iimg\eta_{\mu_1\nu_2}\cdots\eta_{\mu_\spin\nu_\spin}(k_{12})_\rho $};	  

\end{scope}

\end{tikzpicture}
\end{center}
\caption{Three point vertex of two higher spin particles (labelled by index 1 \& 2) and a photon}
\label{fig:krssfeynmanrules2}
\end{figure}

\subsection{Feynman diagrams for four-point functions}
\label{subsec:krsscomptonIfourpointfeyn}

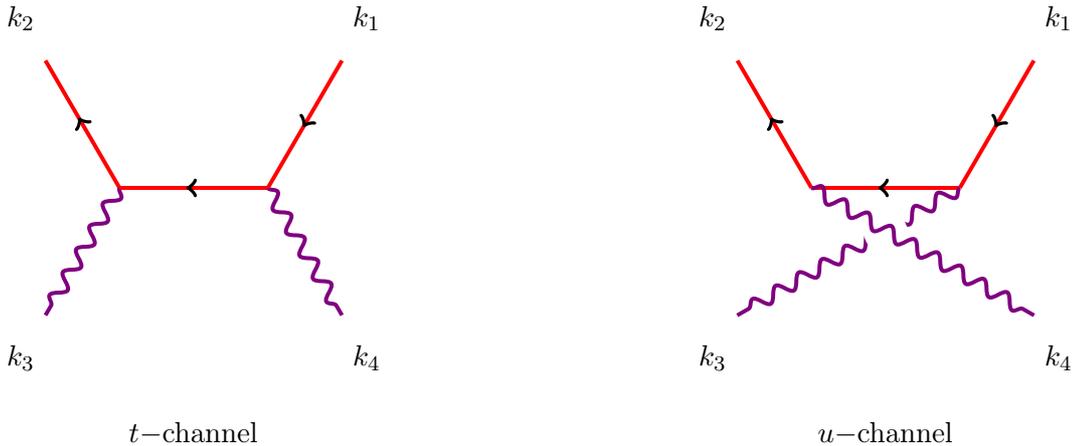
\begin{figure}[h]
\begin{center}
\begin{tikzpicture}[line width=1.5 pt, scale=1.3]
 
\begin{scope}[shift={(0,0)}]	
		
\treefourpointexchange{chspin}{photon}{photon}{chspinbar}{chspin}{0}	

\labeltreefourpointexchange{ $k_2$ }{ $k_3$ }{ $k_4$ }{ $k_1$ }{}{0} 

\draw (0,-2.5) node {$t-$channel};
\end{scope}

\begin{scope}[shift={(7,0)}]	
		
\treefourpointcrossexchange{photon}{chspinbar}{chspin}{photon}{chspin}{180}	

\labeltreefourpointexchange{ $k_2$ }{ $k_3$ }{ $k_4$ }{ $k_1$ }{}{0} 
\draw (0,-2.5) node {$u-$channel};

\end{scope}

\end{tikzpicture} 
\end{center}
\caption{Tree-level Feynman diagrams for the higher spin Compton scattering}
\label{fig:higherspincomptondiagI}
\end{figure}

Let's now briefly discuss the computation of the tree-level four-point function. There are two channels that contribute to this process. There is no $s$-channel diagram in this case\footnote{There would be $s$-channel diagram if photons are replaced by gluons.}. We have depicted the Feynman diagrams in fig. \ref{fig:higherspincomptondiagI}.
In our convention
\begin{enumerate}
	\item All the momenta are outgoing, so the momentum conservation takes the following form  
\begin{equation}
	k_1^\mu +	k_4^\mu +	k_3^\mu +	k_4^\mu =0
\end{equation}
	\item the Mandelstam variables are defined as 
\begin{equation}
	s=(k_1+k_2)^2
\qquad,\qquad	
	t=(k_1+k_4)^2
\qquad,\qquad	
	u=(k_1+k_3)^2
\end{equation}
Even though there are three Mandelstam variables, only two of them are independent
\begin{equation}
	s+t+u= -2m^2
\end{equation}
We choose $t$ and $u$ to be independent. 
\end{enumerate}

\section{A fresh look into of spin-$0$ and spin-$1$}
\label{sec:krsscomptonIscalar}
In this section, we compute Compton scattering for massive spin-$0$ and spin-$1$ particle from the on-shell approach. This will reproduce the known results and convey to the readers the basic philosophy followed in this work. 

\subsection{Spin-0 Compton (Scalar QED)}
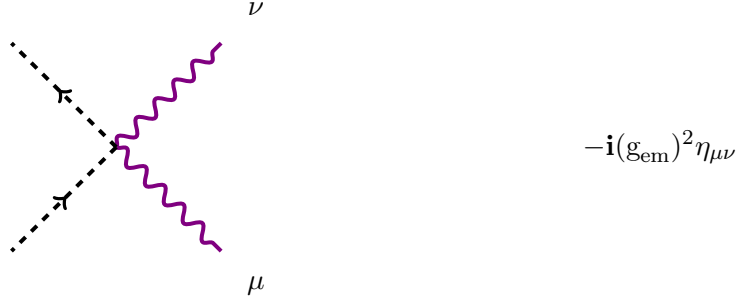
\begin{figure}[h]
\begin{center}
\begin{tikzpicture}[line width=1.5 pt, scale=1.3]

\begin{scope}[shift={(0,5)}]	

\treelevelfourpointskel{photon}{photon}{scalar}{scalarbar}{-45}	
 
\labeltreelevelfourpoint{$\mu $}{$ \nu$}{}{}{-45}
\node at (5.5,0) {$-\iimg  (\gem)^2 \eta_{\mu \nu}$};
 		
\end{scope}	
\end{tikzpicture}
\end{center}
\caption{Two photons interacting with charged scalars (Seagull vertex)} 
\label{fig:scalarqedvertex}
\end{figure}

We begin with the simple case of charged scalar particles\footnote{Usually, we do this computation in scalar QED. However, we follow on-shell methods here. The results derived in this section are not new. However, we present it to orient the readers towards our approach to compute Compton amplitude.}. The three-point function of two scalars and one photon is unique and it is given by 
\begin{equation}
	-\iimg \frac{\gem}{2m} (\varepsilon_3\cdot k_{12})
\label{spinzerocompton1}	
\end{equation} 
With this, we can compute the $t$ and $u$-channel exchange diagrams. We denote the full contribution from the exchange diagrams in the following way
\begin{equation}
		\widetilde{\mathcal{M}}=	\widetilde{\mathcal{M}}^{(t)}+	\widetilde{\mathcal{M}}^{(u)}
\label{spinzerocompton2}	
\end{equation} 
$\widetilde{\mathcal{M}}^{(t)}$ ($\widetilde{\mathcal{M}}^{(u)}$) is the contribution from the $t$-channel ($u$-channel) Feynman diagram; They are given by 
\begin{equation}
\widetilde{\mathcal{M}}^{(t)}=-\iimg \left(\frac{\gem}{m}\right)^2\frac{(\varepsilon_3\cdot k_2)(\varepsilon_4\cdot k_1)}{t+m^2-\iimg \epsilon}	
\qquad,\qquad
\widetilde{\mathcal{M}}^{(u)}=-\iimg\left(\frac{\gem}{m}\right)^2\frac{(\varepsilon_3\cdot k_1)(\varepsilon_4\cdot k_2)}{u+m^2-\iimg \epsilon}	
\label{spinzerocompton3}	
\end{equation}
Let's now check gauge invariance of $\widetilde{\mathcal{M}}$. We begin with the photon in the third leg.  
\begin{equation}
	\widetilde{\mathcal{M}}\Big|_{\varepsilon_3\rightarrow k_3}
	=\frac{1}{2}\iimg\left(\frac{\gem}{m}\right)^2(\varepsilon_4\cdot k_3)
\label{spinzerocompton4}	
\end{equation}
Clearly, the amplitude is not gauge invariant. Moreover, we can see that the RHS is an analytic function of momenta; it has no poles. We can subtract a local four-point vertex $\widetilde{\mathcal{M}}^{(\textrm{ct})}$ to make the gauge invariant. 
\begin{equation}
	\widetilde{\mathcal{M}}^{(\textrm{ct})}=\frac{1}{2}\iimg\left(\frac{\gem}{m}\right)^2(\varepsilon_3\cdot \varepsilon_4)
\label{spinzerocompton5}	
\end{equation}
The contact term is depicted in fig. \ref{fig:scalarqedvertex}. 
We subtract this from $\widetilde{\mathcal{M}}$ in \eqref{spinzerocompton2}.
\begin{equation}
	\widehat{\mathcal{M}}=\widetilde{\mathcal{M}}- \widetilde{\mathcal{M}}^{(\textrm{ct})}
\label{spinzerocompton6}	
\end{equation} 
Clearly, $\widehat{\mathcal{M}}$ vanishes if we substitute pure gauge polarization for the third leg. We can check that it also happens for the fourth leg. 
\begin{equation}
	\widehat{\mathcal{M}}\Big|_{\varepsilon_3\rightarrow k_3}
	=0=\widehat{\mathcal{M}}\Big|_{\varepsilon_4\rightarrow k_4}
\label{spinzerocompton7}	
\end{equation}
Thus, $\widehat{\mathcal{M}}$ gets contribution from the exchange diagrams and the contact term in \eqref{spinzerocompton3} and it is gauge invariant. 

The linearised maxwell field strength is defined as 
\begin{equation}
	\mathcal{B}_{\mu\nu}^{(i)}=k_\mu^{(i)} \epsilon_\nu^{(i)}-k_\nu^{(i)} \epsilon_\mu^{(i)}
\label{raamcomptonbasis1}	
\end{equation}
We can see that the linearised field strength is gauge invariant without using equation of motion (i.e. it is off-shell gauge invariant). 

Since the amplitude in \eqref{spinzerocompton6} is gauge-invariant, it is desirable to write it in terms of the linearised field strength of the massless polarisation; In that case the full amplitude  will be manifestly gauge invariant without using on-shell conditions.

From the linearized Maxwell field strengths (in \eqref{raamcomptonbasis1}), we also define
\begin{equation}
	\mathcal{W}_{\mu \nu}^{(ij)}=\eta^{\rho \sigma}\mathcal{B}_{\mu\rho}^{(i)}\mathcal{B}_{\sigma\nu}^{(j)}	
\end{equation}
In terms of these quantities, the amplitude in \eqref{spinzerocompton6} takes the following form
\begin{equation}
	\widehat{\mathcal{M}}
=
	2\iimg \left(\frac{\gem}{m}\right)^2 \frac{\left( k_1^\mu  \mathcal{W}^{(34)}_{\mu \nu}k_1^\nu\right)}{(t+m^2)(u+m^2)}
\label{spinzerocompton11}	
\end{equation} 
Clearly $	\widehat{\mathcal{M}}$ is off-shell gauge invariant. We see that we need a local four-point vertex to make the amplitude gauge invariant.
 
All of this is very natural in the Lagrangian formalism; all these vertices come from the covariant derivative of the scalar. However, here, we have taken an approach based on amplitudes without relying on Lagrangian, as it can be naturally generalized to higher spin particles. 

\begin{framed}
\noindent
{\bf Lesson 1:}	If we write down on-shell three-point functions and compute the four-point functions due to exchange from these three-point functions, then the four-point exchange may not be gauge invariant. In such a case, we may need to add a local four-point vertex to make the amplitude gauge invariant. 

\end{framed}

\subsection{Compton amplitude of spin-$1$}
\label{sec:krsscomptonIspinone}

We now move to Compton amplitude, which involves massive spin one particles. We denote the polarisations of the massive spin particle 1 with $\zeta_1$ and $\bar \zeta_2$. The three-point function has three kinematically independent structures
\begin{eqnarray}
&&(-\iimg\,)\frac{\gem}{2m} (\zeta_1\cdot \bar\zeta_2)	(\varepsilon_3\cdot k_{12})
\nonumber
\\
&&(-\iimg\,)\frac{\gcou_1}{m} \Big(-(\zeta_1\cdot \varepsilon_3)	(\bar\zeta_2\cdot k_{3}) 
+(\bar\zeta_2\cdot \varepsilon_3)( \zeta_1\cdot k_{3})	\Big)
\label{spinonecompton1}	
\\
&&(-\iimg\,)\frac{\gcou_2}{2m^3} (\zeta_1\cdot k_{3})	(\bar\zeta_2\cdot k_{3})(\varepsilon_3\cdot k_{12})
\nonumber
\end{eqnarray}
Let's start with the simplest case. Let's restrict to $\gcou_1 =0=\gcou_2$ (very soon, we will relax this restriction). The massive spin-$1$ propagator (from \eqref{raamhspinreview5a}) is given by 
\begin{equation}
	\frac{-\iimg }{p^2+m^2-\iimg \epsilon }\left(\eta_{\mu \nu }-\frac{p_\mu p_\nu}{p^2}\right)
\label{spinonecompton2}	
\end{equation}
In this case, the $t$-channel and $u$-channel amplitude turns out to be  
\begin{eqnarray}
\widetilde{\mathcal{M}}^{(t)}&=&
-\iimg \left(\frac{\gem}{m}\right)^2\frac{(\varepsilon_3\cdot k_2)(\varepsilon_4\cdot k_1)}{t+m^2-\iimg \epsilon}
\left((\zeta_1\cdot \bar\zeta_2)+\frac{(\zeta_1\cdot k_4)(\bar\zeta_2\cdot k_3)}{t}\right) 	
\label{spinonecompton3}	
\\
\widetilde{\mathcal{M}}^{(u)}&=&-\iimg \left(\frac{\gem}{m}\right)^2\frac{(\varepsilon_3\cdot k_1)(\varepsilon_4\cdot k_2)}{u+m^2-\iimg \epsilon}\left ((\zeta_1\cdot\bar \zeta_2)+\frac{(\zeta_1\cdot k_3)(\bar\zeta_2\cdot k_4)}{u}\right)
\nonumber
\end{eqnarray}
Note that there is a massless pole in the amplitude; this looks problematic since there is no massless exchange. In order to ascertain whether the pole is physical or not, we should compute the residue at the pole. If the residue vanishes, then it is fine. But, at first, we check gauge invariance! We put pure gauge polarisation for $\varepsilon_3$
\begin{equation}
	\widetilde{\mathcal{M}}\Big|_{\varepsilon_3\rightarrow k_3}
	=\frac{1}{2}\iimg\left(\frac{\gem}{m}\right)^2(\varepsilon_4\cdot k_3)(\zeta_1\cdot\bar\zeta_2)
	- \frac{1}{2}\iimg \left(\frac{\gem}{m}\right)^2\left(\frac{(\varepsilon_4\cdot k_1)(\zeta_1\cdot k_4)(\bar\zeta_2\cdot k_3)}{t}
	+\frac{(\varepsilon_4\cdot k_2)(\zeta_1\cdot k_3)(\bar\zeta_2\cdot k_4)}{u} \right)
\label{spinonecompton4}	
\end{equation}
Just like in the case of spin-$1$, we can see that the amplitude is not gauge invariant. However, in this case, the RHS also has a pole. 

The analytic piece can again be cancelled by a local contact term (to be subtracted) of the following form
\begin{equation}
	\frac{1}{2}\iimg \left(\frac{\gem}{m}\right)^2(\zeta_1\cdot\bar \zeta_2)(\varepsilon_3\cdot \varepsilon_4)
\label{spinonecompton5}	
\end{equation} 
However, the massless pole cannot be cancelled by a local term. It can only be cancelled by a massless exchange. Let's postulate that there is a massless exchange of $\varphi$. We call them ghosts; they are different from the Faddeev-Popov ghosts. They are similar to "would be goldstone" boson(s) in a Higgsed theory or like the ghosts in string theory. We start by postulating a three-point function of the following form 
\begin{equation}
	-\iimg \frac{\textrm{f}_1}{2m^2}(\zeta_1\cdot k_{3})(\varepsilon_3\cdot k_{12})
\label{spinonecompton11}	
\end{equation}

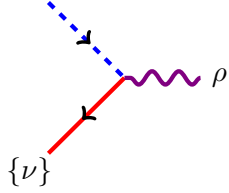
\begin{figure}[h]
\begin{center}
\begin{tikzpicture}[line width=1 pt, scale=1]

\begin{scope}[shift={(0,0)}, rotate=0]


\draw [ultra thick, ghost] (-1,1) -- (0,0);
\draw [ultra thick, chspinbar] (-1,-1) -- (0,0); 
\draw [ultra thick, photon] (1,0) -- (0,0);

\draw (-1.25,-1.25) node {$\{\nu\}$};
\draw (1.25,0) node {$\rho$};


\end{scope}

\end{tikzpicture}
\end{center}
\caption{Ghost vertex}
\label{fig:higherspincomptondiagIIprep}
\end{figure}

\begin{figure}[h]
\begin{center}
\begin{tikzpicture}[line width=1.5 pt, scale=1.3]
 
\begin{scope}[shift={(0,0)}]	
		
\treefourpointexchange{chspin}{photon}{photon}{chspinbar}{ghost}{0}	

\labeltreefourpointexchange{ $k_2$ }{ $k_3$ }{ $k_4$ }{ $k_1$ }{}{0} 

\draw (0,-2.5) node {$t-$channel};
\end{scope}

\begin{scope}[shift={(7,0)}]	
		
\treefourpointcrossexchange{photon}{chspinbar}{chspin}{photon}{ghost}{180}	

\labeltreefourpointexchange{ $k_2$ }{ $k_3$ }{ $k_4$ }{ $k_1$ }{}{0} 
\draw (0,-2.5) node {$u-$channel};

\end{scope}

\end{tikzpicture} 
\end{center}
\caption{Higher spin Compton scattering due to the ghost exchange.}
\label{fig:higherspincomptondiagII}
\end{figure}
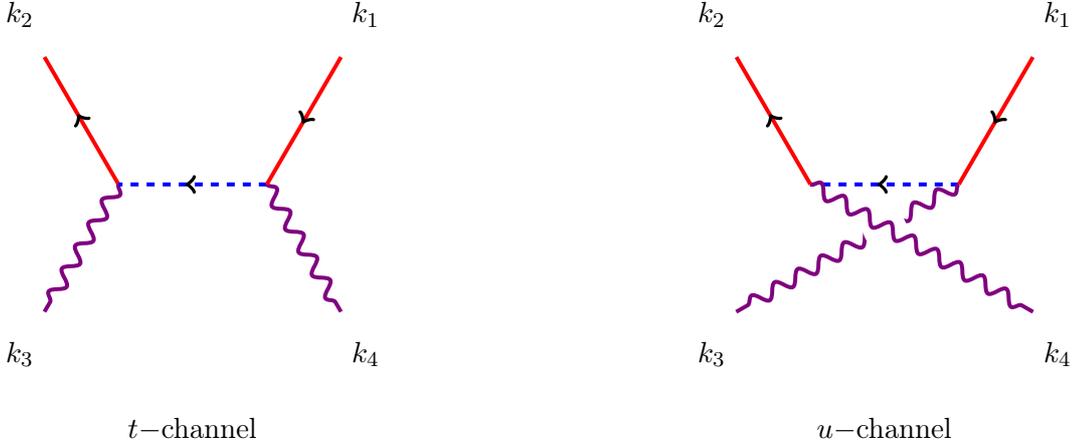
The corresponding Feynman diagram is shown in fig. \eqref{fig:higherspincomptondiagIIprep}; ghosts are denoted in dashed lines. There are two Feynman diagram due to $\varphi$ exchange (see fig. \ref{fig:higherspincomptondiagII}). The sum of the exchanges is given by
\begin{equation}
	-\iimg \frac{\textrm{f}_1^2 }{m^4}\bigg(\frac{(\varepsilon_3\cdot k_2)(\varepsilon_4\cdot k_1)(\zeta_1\cdot k_4)(\bar\zeta_2\cdot k_3)}{t}
	+\frac{(\varepsilon_3\cdot k_1)(\varepsilon_4\cdot k_2)(\zeta_1\cdot k_3)(\bar\zeta_2\cdot k_4)}{u} \bigg)
	\end{equation}
After pure gauge transformation, this reduces to a sum of second term of \eqref{spinonecompton4} plus a local term.
\begin{equation}
\begin{aligned}
	- \iimg \frac{\textrm{f}_1^2}{2m^2}\left(\frac{(\varepsilon_4\cdot k_1)(\zeta_1\cdot k_4)(\bar\zeta_2\cdot k_3)}{t}
	+\frac{(\varepsilon_4\cdot k_2)(\zeta_1\cdot k_3)(\bar\zeta_2\cdot k_4)}{u} \right)\\- \iimg \frac{\textrm{f}_1^2}{2m^4}\bigg((\varepsilon_4\cdot k_1)(\zeta_1\cdot k_4)(\bar\zeta_2\cdot k_3)
	+(\varepsilon_4\cdot k_2)(\zeta_1\cdot k_3)(\bar\zeta_2\cdot k_4)\bigg)
\end{aligned}
\label{spinoneghostamp}
\end{equation}
First term cancels the non-local contribution from \eqref{spinonecompton4} after setting the ghost coupling constant as
\begin{equation} 
	\textrm{f}_1=\pm \iimg \gem 
\label{spinonecompton12}	
\end{equation}
The second part of \eqref{spinoneghostamp} can be countered by a local contact term which gives the same term after pure gauge transformation. Finally the overall contact term (to be subtracted) is given by (we have described the process to find a suitable contact term in section \ref{sec:typeIIcontact}.)
\begin{align}
    &\frac{\iimg \gem^2}{2m^4}\bigg[ {m^2 (\varepsilon_3\cdot \varepsilon_4 )  (\zeta_1\cdot\bar{\zeta}_2 )}+{ (\varepsilon_3  \cdot k_1)  (\zeta_1  \cdot k_3)
        (\varepsilon_4\cdot\bar{\zeta}_2 )}+
        {  (\varepsilon_4 
        \cdot k_1)  (\zeta_1  \cdot k_4)  (\varepsilon_3\cdot\bar{\zeta}_2 )}+
        {  (\varepsilon_3  \cdot k_2)  (\bar{\zeta}_2 \cdot k_3)
        (\varepsilon_4\cdot\zeta_1 )}\nonumber\\ &+{ (\varepsilon_4 
        \cdot k_2) (\bar{\zeta}_2 \cdot k_4)  (\varepsilon_3\cdot\zeta_1 )}
       -\frac{\left(m^2+t\right)  (\varepsilon_3\cdot\bar{\zeta}_2 )  (\varepsilon_4\cdot\zeta_1 )}{2}-\frac{\left(m^2+u\right)  (\varepsilon_3\cdot\zeta_1 )
    (\varepsilon_4\cdot\bar{\zeta}_2 )}{2}\bigg]
        \end{align}

After subtracting the contract term and the contributions from the ghost, the answer is gauge invariant. It can also be written in terms of linearised field strength of the massless polarisation (defined in eqn \eqref{raamcomptonbasis1}).

Then, the whole gauge invariant answer is given by
\begin{equation}
		-\frac{\iimg \gem^2 }{m^2(t+m^2)(u+m^2)}\Bigg[2\left( k_1^\mu  \mathcal{W}^{(34)}_{\mu \nu}k_1^\nu\right)(\zeta_1\cdot \bar\zeta_2) +
\left(\left(k_1\mathcal{B}^{(3)}\zeta_1\right)\left(k_2\mathcal{B}^{(4)}\bar\zeta_2\right)\frac{(t+m^2)}{m^2}+3\leftrightarrow 4 \right)	
		\Bigg]
\label{spinonecompton12.1}	
\end{equation}

\begin{framed}
\noindent
{\bf Lesson 2:}	If we write down on-shell three-point functions and compute the four-point functions due to exchange from these three-point functions, then we already know that the four-point exchange may not be gauge invariant. However, the non-invariant term may have a pole, and in that case, we need to add an unphysical massless exchange (along with a four-point contact term) to ensure gauge invariance. 
\end{framed}

Let's now consider the most general case of parameters in \eqref{spinonecompton1}. We can repeat the analysis, but the answer is not gauge-invariant. The contact term in \eqref{spinonecompton5}  and the exchange due to \eqref{spinonecompton11} is not enough. We need to put one more three-point function
\begin{equation}
	-\iimg\textrm{f}_0(\bar\zeta\cdot  \varphi)
\label{spinonecompton21}	
\end{equation} 
In this case, the spurious poles were cancelled, provided 
\begin{equation}
	 \textrm{f}_0= \pm \iimg \frac{\gcou_1}{2}
\qquad,\qquad 	
\textrm{f}_1=\mp  \iimg \Big(\gcou_1+\gem +\frac{\gcou_2}{2}\Big)
\label{spinonecompton22}	 
\end{equation}
In the standard model of particle physics, $\gem =-\gcou_1$ and $\gcou_2=0$. In that case, $\textrm{f}_1$ turns out to be zero. This also happens to be special in the following sense: the Compton amplitude of spin-$1$ has the ``best"\footnote{All these Compton amplitudes grow for longitudinal polarisations. The growth is least when $\gem =-\gcou_1$ and $\gcou_2=0$.} Ultra-violet behaviour when $\gem = -\gcou_1$ and $\gcou_2=0$. 

\subsubsection{Lesson from the Higgs mechanism}
Readers who are well-versed in the standard model of particles (and/or the Higgs mechanism) will not find it surprising. For the purpose of this discussion, the standard model fermions are irrelevant. In SM, we consider the gauge theory in a Higgsed phase. Consider the Standard model of particle physics. In that case, the $W$ boson becomes massive at low temperature as the vacuum expectation value of the scalar takes a non-zero value. As we increase the temperature, at some point, the vev of the scalar field becomes zero again, and the $W$ boson becomes massless again. This phenomenon is known as electroweak phase transition. The phase in which the $W$ boson has mass is known as the Higgsed phase, and the other phase is known as the unbroken phase. Let's briefly discuss the physics of the Higgs mechanism of $W$ boson, which is a massive spin one particle. In that context, we usually have a scalar field with Mexican hat potential and a gauge theory. When we quantize the theory  in the $R_\xi$ gauge, the $W$ boson propagator turns out to be
\begin{equation}
	\frac{-\iimg }{k^2+m^2-\iimg \epsilon}\left(\eta_{\mu \nu} +\frac{k_\mu k_\nu(\xi-1) }{k^2+\xi m^2-\iimg \epsilon}\right)
\label{spinonecompton31}	
\end{equation} 
The theory also has unphysical particles known as the {\it would-be-goldstone boson}\footnote{There are many ways to see these unphysical particles. In the language of BRST quantisation, they do not belong to BRST cohomology. A simpler (non-rigorous) explanation is that their masses are $\xi$ dependent and thus not gauge invariant.}; they are scalars, and their propagator takes the following form
\begin{equation}
	\frac{-\iimg }{k^2+\xi m^2-\iimg \epsilon} 
\label{spinonecompton32}	
\end{equation} 
Comparing the expression in \eqref{spinonecompton31} with \eqref{raamhspinreview5a}, we can see that we were in the $\xi=0$ gauge. In that case, there is actually an exchange of massless unphysical particles, the {\it would-be} goldstone bosons.

In fact, in the standard model, we can do the computation for any non-negative real values of $\xi$. Any scattering amplitude is independent of $\xi$. $\xi\rightarrow \infty$ is known as the unitary gauge; in that case the ghost decouple. There is one more value of $\xi$, which has interesting features: $\xi=1$. In that case, the propagator has no higher-order poles.

\subsubsection{A re-look at the higher spin propagator}

We can take this clue from the standard model and repeat the computation for an arbitrary value of $\xi$. It turns out that if we choose coupling constants as given in \eqref{spinonecompton22}, then Compton amplitude turns out to be $\xi$ independent. 

Let's define a new quantity as \footnote{We refer the readers to sec V of \cite{Weinberg:1969di}. In our understanding, Weinberg suggested $\xi=\infty$. In \cite{Balasubramanian:2021act}, the authors considered the propagator corresponding to $\xi=\infty$ and demonstrated that the obtained tree-level scattering amplitude which is consistent with partial wave analysis}
\begin{equation}
	\Theta_{\mu \nu}(k,\xi)=\eta_{\mu \nu} +\frac{k_\mu k_\nu(\xi-1) }{k^2+\xi m^2-\iimg \epsilon}
\qquad,\qquad \xi \in [0,\infty)	
\label{comptoncalculation1}	
\end{equation}
From the amplitude perspective, $\xi$ can't be negative to avoid tachyonic particles: From perspectives of Euclidean field theories, it creates a problem with wick rotation. Also tachyonic propagation is inconsistent with causality. 
Let's check the tracelessness and the transversality
\begin{equation}
	\eta^{\mu \nu}\Theta_{\mu \nu}(k,\xi)=d+\frac{k^2(\xi-1) }{k^2+\xi m^2-\iimg \epsilon}
\qquad,\qquad
	k^\mu\Theta_{\mu \nu}(k,\xi)=k_\nu \left(1+\frac{k^2(\xi-1) }{k^2+\xi m^2-\iimg \epsilon} \right)
\label{comptoncalculation2}	
\end{equation}
Let's construct the new projector which is obtained by replacing $\Theta(k)$ with $\Theta(k,\xi)$ in \eqref{raamhspinreview4.1}. 
In the table \ref{tab:xicomparison}, we have provided a comparison of various advantages and disadvantages for the projectors at different values of $\xi$. 

    \begin{table}[!ht]
        \centering
        \begin{tabular}{||p{5cm}|p{1.25cm}|p{1.25cm}|p{1.25cm}|| }
        \hline
        \hline
            ~ & $\xi = 0$ & $\xi=1$ & $\xi=\infty$ \\ \hline
        \hline
           (Off-shell) Tracelessness & \textrm{\color{ao(english)} \cmark} & \textrm{\color{red}\xmark} & \textrm{\color{red}\xmark}  \\ \hline
           (Off-shell)  Transversality & \textrm{\color{ao(english)} \cmark} & \textrm{\color{red}\xmark} & \textrm{\color{red}\xmark} \\ \hline
           Absence of ghost* & \textrm{\color{red}\xmark} & \textrm{\color{red}\xmark} & \textrm{\color{ao(english)} \cmark} \\ \hline
           Absence of Spurious poles & \textrm{\color{red}\xmark} & \textrm{\color{ao(english)} \cmark} & \textrm{\color{ao(english)} \cmark} \\ \hline
            Good UV property & \textrm{\color{ao(english)} \cmark} & \textrm{\color{ao(english)} \cmark} & \textrm{\color{red}\xmark}  \\ \hline
            Absence of IR Divergences & \textrm{\color{red}\xmark} & \textrm{\color{ao(english)} \cmark} & \textrm{\color{ao(english)} \cmark} \\ \hline
           (Off-shell) Idempotence & \textrm{\color{ao(english)} \cmark} & \textrm{\color{ao(english)} \cmark} & \textrm{\color{red}\xmark}  \\ \hline
 \hline
        \end{tabular}
  \caption{Comparison of projector for different values of $\xi$.}
\label{tab:xicomparison}
    \end{table}

We make a few observations:
\begin{itemize}
	\item On-shell, the projector is both transverse and traceless for all $\xi$.

	\item Off-shell the projector is {\it transverse and traceless iff $\xi=0$.}

	\item In the ultraviolet, the propagator diverges badly in the $\xi \rightarrow \infty $ limit, so $\xi\rightarrow 1$ may not be good for loop computations.  In fact this is known for massive spin-$1$ particles in the Lagrangian approach. 
\end{itemize}
Based on the previous discussion, we define a new projector for massive higher spin particles 
\begin{equation}
	\mathcal{P}^{(\spin)}_{(\mu),(\nu)}(p;\xi, m^2 )
\label{comptoncalculation4}	
\end{equation}
which can be obtained from \eqref{raamhspinreview4.1}, simply by replacing all the $ \Theta_{\mu\nu}(p)$s by $\Theta_{\mu \nu}(p,\xi)$. We can always go back to our original propagator by setting $\xi=0$. In the case of spin-$1$ particle, the parameter $\xi$ enters through Faddeev-Popov quantization. We have no such derivation for spin $\spin\geq 2$. 

Let's now move to the ghost propagator. The ghost particles can also be decomposed into irreducible representation of the Poincare group. So we use the same propagator with a small modification. The propagator in \eqref{krsscomptonIresults3} has two pieces, the pole piece and the projector piece. For ghosts, we use the same projector but we change the pole mass to $\xi m^2$
\begin{eqnarray}
		\frac{-\iimg }{k^2+\xi m^2-\iimg \epsilon}	\mathcal{P}^{(\spin)}_{(\mu),(\nu)}(p;\xi, m^2 )
\label{comptoncalculation4.01}	
\end{eqnarray}
For the case of massive spin-$1$ particles, the ghost is spin-$0$ and it's propagator is given by 
\begin{equation}
	\frac{-\iimg }{k^2+\xi m^2-\iimg \epsilon}
\label{comptoncalculation3}	
\end{equation}
The ghost decouples only in the $\xi \rightarrow \infty$ limit. At this point, we note that the string theory answers are available only in the $\xi=1$ gauge; an analogue of the $\xi $ variable is not known in string theory.

 \subsubsection{Simplicity at $\xi=\infty$}
 Let's now consider the case of $\xi=\infty$; in the literature it is known as the {\it Unitary gauge}. This is because the unphysical goldstone becomes infinitely massive and hence decouples from the theory. However, for Higgsed theory for spin-$1$ particles, this gauge is not very friendly for loop computations. Let's consider the projector in \eqref{comptoncalculation1} in the $\xi\rightarrow \infty $ limit (section IV of \cite{Weinberg:1969di})\footnote{We are thankful to R Loganayagam for bringing this reference to our attention and discussions on that.}
 \begin{equation}
	\lim_{\xi\rightarrow \infty }\Theta_{\mu \nu}(k,\xi)=\eta_{\mu \nu} +\frac{k_\mu k_\nu }{ m^2}
\end{equation} 
For a general $\xi$, $\Theta$ has pole at $k^2=-\xi\, m^2$ and thus the projector also has poles. We can see that the projector has no poles in the $\xi\rightarrow \infty$ limit. For spin-$1$ particles, we found that the amplitude is not gauge invariant (see \eqref{spinonecompton4}) and moreover, the gauge non-invariant term also has poles. This poles are essentially coming from the poles in the projector. As a result, if we compute amplitude with $\xi=\infty$ projector and compute gauge invariance of the amplitude, it only has analytic term. In $\xi\rightarrow \infty$ limit, the expression in  \eqref{spinonecompton3} becomes 
			\begin{equation}
					\begin{aligned}
						&\widetilde{\mathcal{M}}_t=-\iimg \bigg(\frac{\gem}{m}\bigg)^2\frac{(\varepsilon_3\cdot k_2)(\varepsilon_4\cdot k_1)}{(t+m^2+\iimg \epsilon)}\bigg((\zeta_1\cdot \bar{\zeta}_2)-\frac{(\zeta_1\cdot k_4)(\bar{\zeta}_2\cdot k_3)}{m^2}\bigg)\\
						&\widetilde{\mathcal{M}}_u=-\iimg \bigg(\frac{\gem}{m}\bigg)^2\frac{(\varepsilon_3\cdot k_1)(\varepsilon_4\cdot k_2)}{(u+m^2+\iimg \epsilon)}\bigg((\zeta_1\cdot \bar{\zeta}_2)-\frac{(\zeta_1\cdot k_3)(\bar{\zeta}_2\cdot k_4)}{m^2}\bigg)	\end{aligned}
			\end{equation}
	After applying the pure gauge transformation $\varepsilon_3\rightarrow k_3$, the sum of $t$ and $u$ channel amplitude transforms as follows
	\begin{equation}
		\frac{\iimg}{2} \bigg(\frac{\gem}{m}\bigg)^2(\varepsilon_4\cdot k_3) (\zeta_1\cdot \bar{\zeta}_2)+	\frac{\iimg}{2m^2} \bigg(\frac{\gem}{m}\bigg)^2\bigg((\varepsilon_4\cdot k_1)(\zeta_1\cdot k_4)(\bar{\zeta}_2\cdot k_3)+(\varepsilon_4\cdot k_2)(\zeta_1\cdot k_3)(\bar{\zeta}_2\cdot k_4)\bigg)
	\end{equation}
Unlike \eqref{spinonecompton4}, the above equation has no poles. It is analytic. We will see that this feature remains true for all higher spins in $\xi\rightarrow\infty$.  Thus the amplitude can be made gauge invariant only by suitable contact interactions. We will be mostly using this simplicity in $\xi=\infty$ in the subsequent sections. 

\subsection{Procedure to find a suitable contact term} 
\label{sec:contacttermsection}
The purpose of this section is to spell out the procedure to find all the gauge non invariant pieces and determine the analytic term whose gauge non invariant piece will cancel the same. There are two types of them, we call them type 1 and type 2. We first show the gauge non-invariant piece when $\varepsilon_i\rightarrow k_i$ for $ i=3,4$, then find the analytic piece which would be required to cancel this gauge non invariance. 

\subsubsection{Type 1}
\label{sec:typeIcont}
First we focus on the gauge non invariant pieces which gives the following type of terms under pure gauge transformations,
\begin{equation}
    \varepsilon_3\rightarrow k_3\quad:\quad \spp{\varepsilon_4} {k_3}\mathcal{F}\qquad,\qquad \varepsilon_4\rightarrow k_4\quad:\quad \spp{\varepsilon_3} {k_4}\mathcal{F}
\end{equation}
Here $\mathcal{F}$ is some function of momentas and massive polarizations (i.e it does not contain any massless polarization).
In this case, the required analytic contact term is simply
    \begin{equation}
        \spp{\varepsilon_3}{\varepsilon_4}\mathcal{F}
    \end{equation}

\subsubsection{Type 2}
\label{sec:typeIIcontact}
Gauge non invariant pieces are,
\begin{equation}
    \varepsilon_3\rightarrow k_3 : (\varepsilon_4\cdot k_1)(\zeta_1\cdot k_4) (\bar \zeta_2\cdot k_3)\mathcal{F}\quad\quad,\qquad  \varepsilon_4\rightarrow k_4 : (\varepsilon_3\cdot k_2)(\zeta_1\cdot k_4) (\bar \zeta_2\cdot k_3)\mathcal{F}
\end{equation}
We have to find a piece involving $\varepsilon_3, \varepsilon_4, \zeta_1$ and $\bar \zeta_2$ in which when $\varepsilon_3 \rightarrow k_3$ gives only the left part, when  $\varepsilon_4 \rightarrow k_4$, gives only the right part. We can start by reverse engineering $k_3 \rightarrow \varepsilon_3$ on the left part resulting in $(\varepsilon_4\cdot k_1)(\zeta_1\cdot k_4) (\bar \zeta_2\cdot \varepsilon_3)\mathcal{F}$ and $k_4 \rightarrow \varepsilon_4$ on the right part resulting in $(\varepsilon_3\cdot k_2)(\zeta_1\cdot \varepsilon_4) (\bar \zeta_2\cdot k_3)\mathcal{F}$. If we add both of them and subtract from the amplitude, then clearly by construction, we can see that the for $\epsilon_3\rightarrow k_3$, the first term cancels the gauge non-invariance of the amplitude; this is by construction. However, in that case, the second term produces an extra gauge invariant piece. Similar thing would happen for gauge transformation of the fourth leg. To summarise, under $\varepsilon_i \rightarrow k_i \ \ i=3,4$, there would be one extra piece left: $$(k_4\cdot k_1)(\zeta_1\cdot k_4) (\bar \zeta_2\cdot \varepsilon_3)\mathcal{F}$$ in case of $i=4$ and $$(k_3\cdot k_2)(\zeta_1\cdot \varepsilon_4) (\bar \zeta_2\cdot k_3)\mathcal{F}$$ in case of $i=3$. 

Thus the search for the contact term is not yet over. We need something more to cancel this extra pieces without creating any more new terms. Noting that, $(k_4\cdot k_1)=(k_3\cdot k_2)=\frac{1}{2}(t+m^2)$, we can reverse engineer both of these terms to single term: $\frac{1}{2}(t+m^2)(\zeta_1\cdot \varepsilon_4) (\bar \zeta_2\cdot \varepsilon_3)\mathcal{F}$, which needs to be subtracted from the previous sum of reverse engineered terms. 

Finally, the resulting required contact term is of the form  
\begin{equation}
   \left( (\varepsilon_4\cdot k_1)(\zeta_1\cdot k_4) (\bar \zeta_2\cdot \varepsilon_3)+ (\varepsilon_3\cdot k_2)(\zeta_1\cdot \varepsilon_4) (\bar \zeta_2\cdot k_3)-\frac{1}{2}(t+m^2)(\zeta_1\cdot \varepsilon_4) (\bar \zeta_2\cdot \varepsilon_3)\right)\mathcal{F}
\end{equation}

\section{Compton amplitude of spin-$2$}
\label{sec:krsscomptonIspintwo}

In this section, we compute the Compton amplitude for spin-$2$; this is the first example where we do not have any help from the Lagrangian formalism.  The case of spin-$2$ has a richer structure than spin-$1$ and spin-$0$. Moreover, it captures all the essential features that repeat for other higher spins as well. As stated above already, we restrict only to the case of the minimal coupling to the photon 
\begin{equation}
	-\iimg \frac{\gem}{2m} (\zeta_1\cdot\bar  \zeta_2)^2(\varepsilon_3\cdot k_{12})
\label{spintwocomptonamplitude1}
\end{equation}
The spin two propagator (after incorporating the modification in \eqref{comptoncalculation4}) is given by 
\begin{equation}
	\frac{-\iimg}{k^2+m^2-\iimg \epsilon }
\mathcal{P}^{(2)}_{\mu_1\mu_2;\nu_1\nu_2}(k,\xi)		
\label{spintwocomptonamplitude2}
\end{equation}
Here the projector $\mathcal{P}^{(2)}_{\mu_1\mu_2;\nu_1\nu_2}(k,\xi)	$ is given by 
\begin{equation}
\mathcal{P}^{(2)}_{\mu_1\mu_2;\nu_1\nu_2}(k,\xi)	
=
\frac{1}{2}\Bigg[
\Theta_{\mu_1\nu_1}(k,\xi)\, 
\Theta_{\mu_2\nu_2}(k,\xi)
+
\Theta_{\mu_1\nu_2}(k,\xi)\, 
\Theta_{\mu_2\nu_1}(k,\xi)
+
2A(2,1,D)
\Theta_{\mu_1\mu_2}(k,\xi)\, 
\Theta_{\nu_1\nu_2}(k,\xi)
\Bigg] 	
\label{spintwocomptonamplitude3}
\end{equation}
$A(2,1,D)$ is given by $-\frac{1}{D-1}$. Then, the $t$-channel amplitude is given by 
\begin{equation}
 \begin{aligned}
\widetilde{\mathcal{M}}^{(t)}(\xi)=&-\iimg \left(\frac{\gem}{m}\right)^2\frac{(\varepsilon_4\cdot k_1)(\varepsilon_3\cdot k_2)}{t+m^2-\iimg \epsilon}\left [(\zeta_1\cdot \bar \zeta_2)^2-2(\xi-1)(\zeta_1\cdot \bar \zeta_2)\frac{(\zeta_1\cdot k_4)(\bar \zeta_2\cdot k_3)}{t+\xi m^2}\right.
\\
&\left.
+
\frac{D-2}{D-1}
(\xi-1)^2\frac{(\zeta_1\cdot k_4)^2(\bar \zeta_2\cdot k_3)^2}{(t+\xi m^2)^2}
\right]
\label{spintwocomptonamplitude4}
  \end{aligned}
\end{equation}
The $u$-channel amplitude can obtained simply by $3\leftrightarrow 4$ exchange
\begin{equation}
\widetilde{\mathcal{M}}^{(u)}(\xi)=\widetilde{\mathcal{M}}^{(t)}(\xi)\Bigg|_{3\leftrightarrow 4}	
\label{spintwocomptonamplitude5}
\end{equation}
The amplitude, obtained from adding the $t$-channel and $u$-channel exchange, is given by 
\begin{equation}
\widetilde{\mathcal{M}}(\xi)=\widetilde{\mathcal{M}}^{(t)}(\xi)+\widetilde{\mathcal{M}}^{(u)}(\xi)	
\label{spintwocomptonamplitude11}
\end{equation}
Now, we check the gauge invariance of the amplitude. Let's check it for the third leg first. 
\begin{equation}
 \begin{aligned}
	\widetilde{\mathcal{M}}(\xi)\Bigg|_{\varepsilon_3\rightarrow k_3}
&=-\iimg \left(\frac{\gem}{m}\right)^2(\varepsilon_3\cdot k_2)\left [(\zeta_1\cdot \bar \zeta_2)^2-2(\xi-1)(\zeta_1\cdot \bar \zeta_2)\frac{(\zeta_1\cdot k_4)(\bar \zeta_2\cdot k_3)}{t+\xi m^2}
\right.
\\
&\left.+
\frac{D-2}{D-1}
(\xi-1)^2\frac{(\zeta_1\cdot k_4)^2(\bar \zeta_2\cdot k_3)^2}{(t+\xi m^2)^2}
\right]	+1\leftrightarrow 2
    \end{aligned}
\label{spintwocomptonamplitude12}
\end{equation}
We can see that the RHS have both analytic and non-analytic pieces. It is very similar to spin-$1$. We follow the same strategy. We add contact interaction and particles of lower spin to restore gauge invariance. In this case, the unphysical particles can have spin-$0$ or 1. Let's consider the tensor structure for the third term in the above expression 
\begin{equation}
 \begin{aligned}
(\zeta_1\cdot k_4)^2(\bar \zeta_2\cdot k_3)^2
    \end{aligned}
\label{spintwocomptonamplitude13}
\end{equation}
This tensor structure cannot appear due to spin-$0$ or spin-$1$ exchange  (See sec \ref{sec:krsscomptonIunitarity} for details). As a result, it is not possible to cancel this term by adding a particle of spin-$0$ or spin-$1$. This demonstrates the point that the amplitude is not gauge invariant unless $\xi=1$ (in this case, the coefficient of the term vanishes and thus the term is not present) or $\xi=\infty$ (in this case, the pole is absent, and thus we hope to restore gauge invariance by adding a contact interaction). In particular, it is not possible to restore gauge invariance for $\xi=0$ (i.e. for the original propagator given in \eqref{raamhspinreview5a}). So, we first consider the case of $\xi=\infty$ as it is simpler to analyze.

\subsection{Unitary gauge ($\xi=\infty$)}
\label{sec:krsscomptonIspintwoug}

We consider \eqref{spintwocomptonamplitude12} and take the limit $\xi\rightarrow \infty$. In this case, the RHS is completely analytic. So, it is possible to make it gauge invariant following the procedure in sec \ref{sec:contacttermsection}. The relevant contact term is 
\begin{equation}
		\begin{aligned}
			& \frac{\iimg (\gem)^2}{2m^2}\Bigg[ (\zeta_1\cdot\bar \zeta_2)^2(\varepsilon_3\cdot\varepsilon_4)+\frac{1}{m^2}
			\bigg\{
			(\zeta_1\cdot k_4) (\zeta_2\cdot \varepsilon_3)(\varepsilon_4\cdot k_1)+\spp{\zeta_2}{k_3}\spp{\zeta_1}{\varepsilon_4}\spp{\varepsilon_3}{k_2}\nonumber\\
			&-\frac{(t+m^2)}{2}\spp{\zeta_1}{\varepsilon_4}\spp{\zeta_2}{\varepsilon_3}
			\bigg\}\bigg\{2(\zeta_1\cdot\bar\zeta_2)-\frac{1}{m^2}\left(\frac{D-2}{D-1}\right)(\zeta_1\cdot k_4)( \bar \zeta_2\cdot k_3)\bigg\}+3\leftrightarrow 4 \Bigg]
		\end{aligned}
\end{equation}
After adding the contact interaction, the amplitude is gauge invariant. We can write the amplitude on the gauge invariant basis. 
{\begin{equation}
		\begin{aligned}
			&\frac{\iimg (\gem)^2}{m^2(t+m^2)(u+m^2)}\Bigg[2 \left( k_1^\mu  \mathcal{W}^{(34)}_{\mu \nu}k_1^\nu\right)(\zeta_1\cdot\bar\zeta_2)^2
			\\        
			&+\Bigg\{\frac{(u+m^2)}{m^2} \left(k_1\mathcal{B}^{(4)}\zeta_1\right)\left(k_2\mathcal{B}^{(3)}\bar\zeta_2\right)\Bigg(2(\zeta_1\cdot\bar\zeta_2)-\frac{1}{m^2}\left(\frac{D-2}{D-1}\right)(\zeta_1\cdot k_4)( \bar \zeta_2\cdot k_3)\Bigg)+3\leftrightarrow 4\Bigg\}\Bigg]
		\end{aligned}
\end{equation}

\subsection{Feynman gauge ($\xi=1$)}
\label{sec:krsscomptonIspintwofg}

Let's now consider the Feynman gauge (i.e. $\xi=1$). In that case,  the RHS in \eqref{spintwocomptonamplitude12} also has no poles. From the spin-$1$ example, we already know that there are unphysical goldstone bosons, and their exchange cancels the propagation of unphysical (time-like) d.o.f of the massive particle to ensure unitarity. The same thing happens here. In this case, we need to consider two ghosts\footnote{We use the word "ghost" in the sense that they are not physical particles; however, they are different from Faddeev-Popov ghosts}: one with spin one and another spin zero. We consider the following three-point function (figure \ref{fig:higherspincomptondiagIIprep})
\begin{eqnarray}
	(-\iimg)  \left(\frac{\gcou_{\texttt{j}}^{(2)} }{2m^{3-\texttt{j}}}\right) (\zeta_1\cdot  \varphi_2)^{\texttt{j}}(\bar \zeta_2\cdot k_{3})^{2-\texttt{j}} (\varepsilon_3\cdot k_{12})
\qquad,\qquad
\texttt{j}=0,1	
\label{comptonminimalcalculation3}
\end{eqnarray}
We can compute the Compton amplitude (depicted in figure \ref{fig:higherspincomptondiagII}) due to these ghost exchanges and it restores unitarity provided 
\begin{equation}
    \gcou_0^{(2)} = \pm\frac{\gem }{\sqrt{2}}\qquad;\qquad \gcou_1^{(2)}=\pm\frac{\iimg\gem }{4}\sqrt{\frac{D-2}{D-1}}
\end{equation}
Unlike spin-$1$, the addition of these unphysical particles cannot restore unitarity for any finite value of $\xi\ne 1$.
\begin{framed}
\noindent
{\bf Lesson 3:}	From the spin two computations, we have learnt that it is possible to obtain unitary Compton amplitude only for $\xi=1$ and $\xi=\infty$. In particular, if we consider the propagator for $\xi=0$, then the propagator is off-shell transverse and traceless, but it is not possible to get a gauge-invariant amplitude. 
\end{framed}

\section{Compton amplitude of arbitrary spin}
\label{sec:krsscomptonIarbitraryspin}

We have already done a few cases. Now, we present the results for all spins. For the purpose of simplicity, we restrict to only minimal coupling. 
\begin{equation}
	(-\iimg\,)\frac{\gem}{2m} (\zeta_1\cdot \bar \zeta_2)^\spin 	(\varepsilon_3\cdot k_{12})
\label{comptonminimalcalculation1}
\end{equation}
We want to compute the tree-level Compton amplitude due to this three-point function. There are two diagrams that contribute to the process (see fig. \ref{fig:higherspincomptondiagI}). 
We start by computing the $t$-channel amplitude. 
%
We write it as
\begin{equation}
\widetilde{\mathcal{M}}^{(t)}=
\frac{-\iimg}{t+m^2-\iimg \varepsilon}
\left(\frac{\gem}{m}\right)^2(\varepsilon_4\cdot k_1)(\varepsilon_3\cdot k_2) \zeta_1^{\{\mu\}} \mathcal{P}^{(\spin)}_{\{\mu\},\{\nu\}}(k_1+k_4, \xi)\bar \zeta_2^{\{\nu\}} 
\label{comptonminimalcalculation13}
\end{equation}
$\mathcal{P}^{(\spin)}$ is defined in \eqref{comptoncalculation4}, which is essentially a modified form of \eqref{raamhspinreview4.1}. Putting that expression, we get
\begin{eqnarray}
-\iimg\left(\frac{\gem}{m}\right)^2\frac{(\varepsilon_4\cdot k_1)(\varepsilon_3\cdot k_2)}{t+m^2-\iimg \varepsilon}\Bigg[\sum_{a=0}^{\lfloor \spin/2\rfloor}
    A(\spin,a,D)\, (\zeta_1\odot_{t} \zeta_1)^{a}(\bar \zeta_2\odot_{t} \bar \zeta_2)^{a}(\zeta_1\odot_{t}\bar \zeta_2)^{\spin-2a}\Bigg]
\label{comptonminimalcalculation14}
\end{eqnarray}
 $\odot$ is the dot product defined with respect to $\Theta_{\mu \nu }$; definition of $\Theta$ depends on momenta. $\odot_t$ denotes the fact that internal momenta is $k_1^\mu+k_4^\mu$. The expressions for the various $\theta$ dot product appearing in \eqref{comptonminimalcalculation14} are given by 
\begin{eqnarray}
(\zeta_1\odot_{t}\bar \zeta_1)&=&\frac{(\xi-1)}{t+\xi m^2-\iimg \varepsilon}(\zeta_1\cdot k_4)^2
\label{comptonminimalcalculation15.1}
\\	
(\bar \zeta_2\odot_{t}\bar \zeta_2)&=&\frac{(\xi-1)}{t+\xi m^2-\iimg \varepsilon}( \bar \zeta_2\cdot k_3)^2
\label{comptonminimalcalculation15.2} 
\\	
(\zeta_1\odot_{t} \bar \zeta_2)&=&\left(\zeta_1\cdot \bar \zeta_2-\frac{(\xi-1)}{t+\xi m^2-\iimg \varepsilon}( \zeta_1\cdot k_4)(  \bar \zeta_2\cdot k_3)\right)
\label{comptonminimalcalculation15.3}
\end{eqnarray}
Putting these expression back into \eqref{comptonminimalcalculation14}, we get 
\begin{multline}
\label{comptonminimalcalculation21}
 -\iimg\frac{(\gem)^2}{m^2} \frac{(\varepsilon_4\cdot k_1)(\varepsilon_3\cdot k_2)}{t+m^2-\iimg\varepsilon}  \\
    \times \sum\limits_{a=0}^{\lfloor\frac{\spin}{2}\rfloor} A(\spin,a,D) \left\{\frac{(\xi-1)(\zeta_1\cdot k_4)(\bar \zeta_2\cdot k_3)}{t+\xi m^2-\iimg \varepsilon}\right\}^{2a} 
    \left\{(\zeta_1\cdot \bar \zeta_2) - \frac{(\xi-1)(\zeta_1\cdot k_4)(\bar \zeta_2\cdot k_3)}{t+\xi m^2-\iimg \varepsilon}\right\}^{\spin-2a}
\end{multline}
Let us focus on the summation in the second line. To write the expression in a linear combination of amplitude basis form, we binomial expand the second term in curly braces. 
\begin{multline}
 -\iimg  \frac{\gem^2}{m^2}  \frac{(\varepsilon_4\cdot k_1)(\varepsilon_3\cdot k_2)}{t+m^2-\iimg\varepsilon} 
    \times \sum\limits_{a=0}^{\lfloor\frac{\spin}{2}\rfloor} A(\spin,a,D) \sum\limits_{b=0}^{\spin-2a} {}^{\spin-2a}\mathrm{C}_b\  (\zeta_1\cdot \bar \zeta_2)^b  \left(-\frac{(\xi-1)(\zeta_1\cdot k_4)(\bar \zeta_2\cdot k_3)}{t+\xi m^2-\iimg \varepsilon}\right)^{\spin-b}
\label{comptonminimalcalculation23}
\end{multline}
Now we want to interchange the two summations in the above formula.

For future purposes, it is very convenient to define a new quantity as following 
\begin{equation}
 \label{comptonminimalcalculation24}
    B(\spin,b,D) = \sum\limits_{a=0}^{\lfloor \frac{\spin-b}{2} \rfloor}  {}^{\spin-2a}\mathrm{C}_b \ A(\spin,a,D)(-1)^{\spin-b}
\end{equation}
With the help of this $ B(\spin,b,D) $ we can interchange the summations in \eqref{comptonminimalcalculation23} and write it as 
\begin{multline}
\label{comptonminimalcalculation25}
 -\iimg \frac{(\gem)^2}{m^2} (\varepsilon_3\cdot k_2)(\varepsilon_4\cdot k_1)
         \left[ \sum\limits_{b=0}^{\spin} B(\spin,b,D) (\zeta_1\cdot \bar \zeta_2)^b (\zeta_1\cdot k_4)^{\spin-b} (\bar \zeta_2\cdot k_3)^{\spin-b}
                \left(\frac{\xi-1}{t+\xi m^2-\iimg \varepsilon}\right)^{\spin-b}   \right]
\end{multline}
In the above expression, we separately write the $b=\spin$ piece. The motivation behind this will become clear in the next section 
\begin{multline}
\label{comptonminimalcalculation26}
 -\iimg \frac{(\gem)^2}{m^2} (\varepsilon_3\cdot k_2)(\varepsilon_4\cdot k_1)
         \left[  (\zeta_1\cdot \bar \zeta_2)^\spin 
                +\sum\limits_{b=0}^{\spin-1} B(\spin,b,D) (\zeta_1\cdot \bar \zeta_2)^b (\zeta_1\cdot k_4)^{\spin-b} (\bar \zeta_2\cdot k_3)^{\spin-b}
                \left(\frac{\xi-1}{t+\xi m^2-\iimg \varepsilon}\right)^{\spin-b}   \right]
\end{multline}
In the case of spin-$2$, we have already seen above that the analysis becomes simple for $\xi=\infty$. So, we consider that case first. 

\subsection{Unitary gauge ($\xi=\infty$)}
\label{sec:krsscomptonIarbitraryspinUnitarygauge}

When we take the $\xi \rightarrow \infty$ limit, the expression, given in \eqref{comptonminimalcalculation15.1}, \eqref{comptonminimalcalculation15.2} \& \eqref{comptonminimalcalculation15.3}, simplifies due to the following reason 
\begin{equation}
	\lim_{\xi \rightarrow \infty}\frac{(\xi-1)}{t+\xi m^2-\iimg \varepsilon}
	= \frac{1}{m^2}
 \label{comptonminimalcalculation41}
\end{equation} 
This also implies that all the higher-order poles disappear. In this limit, \eqref{comptonminimalcalculation26} becomes
\begin{multline}
 -\iimg\frac{(\gem)^2}{m^2} (\varepsilon_3\cdot k_2)(\varepsilon_4\cdot k_1)
    \left[(\zeta_1\cdot \bar \zeta_2)^\spin
    + \frac{(\zeta_1\cdot k_4) (\bar \zeta_2\cdot k_3)}{m^2}  \sum\limits_{b=0}^{\spin-1} B(\spin,b,D)\  (\zeta_1\cdot \bar \zeta_2)^b\  \left\{\frac{(\zeta_1\cdot k_4)(\bar \zeta_2\cdot k_3)}{m^2}\right\}^{\spin-1-b}\right]
 \label{comptonminimalcalculation42}
\end{multline}
Now, we check the gauge invariance of the amplitude. To make further calculations simpler and easier to comprehend, we introduce two new quantities:
\begin{eqnarray}
\Omega(\spin)&=&(\zeta_1\cdot \bar \zeta_2)^\spin
\\
    \Upsilon^{(t)}(\spin) &=&  \sum\limits_{b=0}^{\spin-1} B(\spin,b,D)\  (\zeta_1\cdot \bar \zeta_2)^b\
    \left(\frac{(\zeta_1\cdot k_4)(\bar \zeta_2\cdot k_3)}{m^2}\right)^{\spin-b-1}
 \label{comptonminimalcalculation43}
\end{eqnarray}
With this notation, the $t$-channel amplitude becomes
\begin{equation}
	\widetilde{\mathcal{M}}^{(t)}=
	-\iimg  \frac{(\gem)^2}{m^2} \frac{(\varepsilon_3\cdot k_2)(\varepsilon_4\cdot k_1)}{t+m^2-\iimg  \epsilon} 
    \left[\Omega(\spin)
    + \frac{(\zeta_1\cdot k_4) (\bar \zeta_2\cdot k_3)}{m^2} \Upsilon^{(t)}(\spin)\right]
 \label{comptonminimalcalculation44}
\end{equation}
The $u$-channel amplitude can obtained simply by $3\longleftrightarrow 4$ exchange.
The whole amplitude, obtained from adding the $t$-channel and $u$-channel exchange, is given by 
\begin{equation}
\widetilde{\mathcal{M}}=\widetilde{\mathcal{M}}^{(t)}+\widetilde{\mathcal{M}}^{(u)}
 \label{comptonminimalcalculation51}
\end{equation}
Let's first check gauge invariance for 3rd leg 
\begin{align}
    \widetilde{\mathcal{M}}\bigg\vert_{\varepsilon_3\to k_3} = -\frac{1}{2}\iimg \left(\frac{\gem}{m}\right)^2 
    &(\varepsilon_4\cdot k_1) \left(\Omega(\spin)+\frac{(\zeta_1\cdot k_4) (\bar \zeta_2\cdot k_3)}{m^2} \Upsilon^{(t)}(\spin)\right) \nonumber\\
&-\frac{1}{2}\iimg\left(\frac{\gem}{m}\right)^2 \left[
    (\varepsilon_4\cdot k_2)\left(\Omega(\spin)+\frac{(\bar \zeta_2\cdot k_4) (\zeta_1\cdot k_3)}{m^2} \Upsilon^{(u)}(\spin)  \right)
    \right]\nonumber
\end{align}
\begin{align}
\label{comptonminimalcalculation52}
    = -\frac{1}{2}\iimg\left(\frac{\gem}{m}\right)^2 \left[
        -(\varepsilon_4\cdot k_3)\Omega(\spin)+ \frac{(\varepsilon_4\cdot k_1)(\zeta_1\cdot k_4) (\bar \zeta_2\cdot k_3)}{m^2} \Upsilon^{(t)}(\spin)+\frac{(\varepsilon_4\cdot k_2)(\bar \zeta_2\cdot k_4) (\zeta_1\cdot k_3)}{m^2} \Upsilon^{(u)}(\spin)
    \right]
\end{align}
Clearly, the amplitude is not gauge invariant. However, though not unexpected, it does not have any poles. The expression is completely analytic and hence may be removed by adding a local contact interaction. In fact, after this special arrangement in terms of $\Omega$ and $\Upsilon$s, it becomes easier to determine the contact term required for the gauge invariance. This is because, we have pulled out the dependence on $\varepsilon$s outside the $\Omega$ and $\Upsilon$ pieces, so whatever gauge noninvariance is seen, is present only in the coefficients of $\Omega$ and $\Upsilon$. We can also determine the contact terms keeping $\Omega$ and $\Upsilon$s as it is but changing their coefficients. We can do the same in \eqref{comptonminimalcalculation52} by using results from sec \ref{sec:typeIcont} on coefficient of $\Omega$ and \ref{sec:typeIIcontact} on coeffcient of $\Upsilon^{(t)}$($3\leftrightarrow 4$ of \ref{sec:typeIIcontact} for coefficient of $\Upsilon^{(u)}$). Finally, after all this, the contact term is given by, 

\begin{equation}
\label{comptonminimalcalculation53}
		\begin{aligned}
			& \frac{\iimg (\gem)^2}{2m^2}\Bigg[ (\zeta_1\cdot\bar \zeta_2)^\spin(\varepsilon_3\cdot\varepsilon_4)-\frac{1}{m^2}
			\bigg\{
			(\zeta_1\cdot k_4) (\zeta_2\cdot \varepsilon_3)(\varepsilon_4\cdot k_1)+\spp{\zeta_2}{k_3}\spp{\zeta_1}{\varepsilon_4}\spp{\varepsilon_3}{k_2}\nonumber\\
			&-\frac{(t+m^2)}{2}\spp{\zeta_1}{\varepsilon_4}\spp{\zeta_2}{\varepsilon_3}
			\bigg\}\Upsilon^{(t)}(\spin)+3\leftrightarrow 4 \Bigg]
		\end{aligned}
\end{equation}
The total gauge-invariant amplitude is obtained from adding \eqref{comptonminimalcalculation51} and \eqref{comptonminimalcalculation53}. We can write using a gauge-invariant basis. The amplitude turns out to be 
\begin{equation}
		\begin{aligned}
			&\frac{\iimg (\gem)^2}{m^2(t+m^2)(u+m^2)}\Bigg[2 \left( k_1^\mu  \mathcal{W}^{(34)}_{\mu \nu}k_1^\nu\right)(\zeta_1\cdot\bar\zeta_2)^2
		-\Bigg\{\frac{(u+m^2)}{m^2} \left(k_1\mathcal{B}^{(4)}\zeta_1\right)\left(k_2\mathcal{B}^{(3)}\bar\zeta_2\right)\Upsilon^{(t)}(\spin)+3\leftrightarrow 4\Bigg\}\Bigg]
		\end{aligned}
\end{equation}

\subsection{Feynman gauge ($\xi=1$)}
\label{sec:krsscomptonIarbitraryspinFeynmangauge}

We now discuss the Compton amplitude for $\xi=1$. We need ghosts coupling with the higher spin field at $\xi=1$ to make the amplitude unitary; we have shown this in sec \ref{subsubsec:krsscomptonIunitarityVII}. But how can we calculate the ghost couplings? The idea is that the ghosts would decouple at $\xi=\infty$. So if we take the difference of amplitude at $\xi=\infty$ and $\xi=1$, whatever is left should be coming due to the ghost exchange, which we need to add to the $\xi=1$ case to make the amplitude unitary. The proof of this statement can be found in sec \ref{subsubsec:krsscomptonIunitarityVII}.

Let us take a look at the difference,
\begin{equation}
    \widetilde{\mathcal{M}}^{(t)}\bigg\vert_{\xi\to\infty} - \widetilde{\mathcal{M}}^{(t)}\bigg\vert_{\xi\to 1} = 
    -\iimg  \frac{(\gem)^2}{m^2} \frac{(\varepsilon_3\cdot k_2)(\varepsilon_4\cdot k_1)}{t+m^2-\iimg \varepsilon}
    \sum\limits_{b=0}^{\spin-1} B(\spin,b,D)(\zeta_1\cdot \bar \zeta_2)^b\  \left(\frac{(\zeta_1\cdot k_4)(\bar \zeta_2\cdot k_3)}{m^2}\right)^{\spin-b}
\end{equation}
Inspired by the ghost propagators from the electroweak sector of the standard model, we propose the propagator of spin-$\texttt{j}$ ghost to be,
\begin{equation}
  \frac{-\iimg}{p^2+\xi m^2-\iimg\epsilon} \mathcal{P}^{(\texttt{j})}_{\{\mu\}\{\nu\}}(p, \xi)
\end{equation}
The corresponding Feynman diagrams are schematically of the same form as in fig \ref{fig:higherspincomptondiagII}. The only difference is that in this case we need to add a tower of ghosts with spin ranging from $\spin-1$ to $0$ with a precise coupling given by 
\begin{eqnarray}
	-\iimg g^{(\spin)}_{\texttt{j}} \left(\frac{\gem}{2m^{\spin-\texttt{j}+1}}\right) (\zeta_1\cdot  \varphi_2)^\texttt{j}(\bar \zeta_1\cdot k_{3})^{\spin-\texttt{j}} (\varepsilon_3\cdot k_{12})
\qquad,\qquad
	0\leq \texttt{j}< \spin 
\end{eqnarray}
and the coupling constant turns out to be 
\begin{equation}
  g^{(\spin)}_{\texttt{j}} =  \pm  \iimg \sqrt{B(\spin,\texttt{j},D)}   
\end{equation}
It is not possible to construct any unitary amplitude for any finite $\xi\ne 1$. We show the proof for this in sec \ref{subsubsec:krsscomptonIunitarityVII}.

\section{Unitarity and cutting rules}
\label{sec:krsscomptonIunitarity}

In this section, we discuss the unitarity of QFT and what it implies for Compton amplitude. The readers can check, for example,  \cite{tHooft:1973wag} for an introduction to Unitarity in Quantum Field Theory. 

\subsection{Cutkosky rules}
\label{subsec:krsscomptonIunitarityI}

The total probability in a quantum theory should always be one, and it should remain the same under time evolution. This means the time evolution $\mathcal{U}(t_1,t_2)$ operator has to be unitary. $S$ matrix is defined to be 
\begin{equation}
	S=\mathcal{U}(\infty,-\infty)
\label{krsunitarity1}	
\end{equation}
This implies the $S$ satisfies the following property\footnote{Let $\mathcal{H}_\invc$ ($\mathcal{H}_\outvc$) be the Hilbert space formed out of $|\invc\rangle$ ($|\outvc\rangle$) states and $H_{\textrm{Free}}$ is the free Hamiltonian. $S$ is a map from $\mathcal{H}_\invc$ to $\mathcal{H}_\outvc$. This in a most general situation $ S^\dagger S=\mathrm{\textbf{1}}_{\invc}$ and $SS^{\dagger} $ is $ S^\dagger S=\mathrm{\textbf{1}}_{\outvc}$. We are working with the assumption that $|\invc\rangle$ is isomorphic to $\mathcal{H}_\outvc$. 
}
\begin{equation}
	SS^{\dagger} = S^\dagger S = \mathrm{\textbf{1}}
\label{krsunitarity3}	
\end{equation}
$S$ matrix includes all sorts of processes. Usually, the $S$ matrix is written in the following way 
\begin{equation}
	S = \mathrm{\textbf{1}} + T
\label{krsunitarity5}	
\end{equation}
$\mathrm{\textbf{1}} $ denotes the free evolution; $T$ denotes all the possible evolutions due to interactions in the theory. Then, the unitarity condition becomes
\begin{align}
	(T + T^\dagger) &= - T T^\dagger
\label{krsunitarity7}	
\end{align}
Note that on the left, we have terms with a single factor of $T$; on the right, we have terms with two factors.

In a Perturbative Quantum Field Theory, we compute $T$  order by order in perturbation theory. In this paper, the perturbative parameter is $\gem$\footnote{More precisely, the perturbative parameter is also a dimensionless number formed out of $\gem$, $\hbar$, $m$ and other dimensionless numbers. For example, in $3+1$ QED, the perturbative parameter is $\gem/\sqrt{4\pi}$.  }; let the perturbative expansion of $T$ matrix is given by 
\begin{equation}
	T=\sum_{n=0}^\infty(\gem)^n\,  T^{(n)}\qquad,\qquad T^{(0)}=0
\label{krsunitarity9}	
\end{equation}
The second equation implies that the transfer matrix is zero in the free theory. If we put this in \eqref{krsunitarity7}, and equate terms which are same order in $\gem$, we get 
\begin{equation}
		T^{(n)} + (T^{(n)})^\dagger = - \sum_{n_1=0}^{n} T^{(n)}(T^{(n-n_1)})^\dagger
\label{krsunitarity11}	
\end{equation}
Thus, in a perturbative QFT, unitarity implies a relation between contributions at different orders in perturbation. 

Eqn \eqref{krsunitarity7} gives the relation in terms of operators. $S$ matrix elements are different matrix elements of this operator. We check a particular matrix element of the operator by sandwiching it in between out($\langle \outvc |$) and in($|\invc\rangle$) states,
\begin{align}
	\langle \outvc|T|\invc\rangle + \langle \invc | T | \outvc \rangle ^* &= - \langle \outvc | T T^\dagger | \invc \rangle
\label{krsunitarity13}	
\end{align} 
Now, we want to put completeness relation in the RHS. The completeness relation in QFT takes the following form 
\begin{equation}
	\mathrm{\textbf{1}}= \int \mathcal{D\sigma}\  |\sigma \rangle\langle \sigma |=\sum_N\int \mathcal{D\sigma}_N\  |\sigma_n \rangle\langle \sigma_n |
\label{krsunitarity15}	
\end{equation} 
$N$ is the number of particles. More explicitly, the terms in the sum are given by 
\begin{equation}
\int \mathcal{D\sigma}_N\  |\sigma_N \rangle\langle \sigma_N |
=\int \prod_{i=0}^{N}\frac{d^{D-1} q_i}{(2\pi)^{D-1}}\
	\frac{1}{2E_q} |\{q_N\} \rangle\langle \{q_N\} | 
\label{krsunitarity15}	
\end{equation}
Now we put this completeness relation in \eqref{krsunitarity13}
to obtain
\begin{align}
- \sum_N\int \mathcal{D\sigma}_N \langle \outvc | T |\sigma_N \rangle\langle \sigma_N |T^\dagger | \invc \rangle = - \int \mathcal{D\sigma}_N \langle \outvc | T |\sigma_N \rangle (\langle \invc |T | \sigma_N \rangle)^*
\label{krsunitarity17}	
\end{align} 
Let us write the above expression in terms of the Scattering amplitudes. In scattering amplitude, we write the momentum conservation part as a separate delta function 
\begin{equation}
    \langle f | T | i \rangle =  (2\pi)^4 \delta^4(p_i + p_f) \mathcal{M}(i\rightarrow f)
\label{krsunitarity19}	
\end{equation} 
here $|i\rangle$ and $|f\rangle$ denote the initial and final states, respectively. Inside the delta function, $p_i$ and $p_f$ denote the sum of momenta of all initial and final particles, respectively. Note that we have been calculating this $\mathcal{M}$ till the last section. It would be convenient to write the unitarity relation in terms of this quantity. Writing \eqref{krsunitarity13} in terms of $\mathcal{M}$, we get,
\begin{equation}
    \mathcal{M}(\invc \rightarrow \outvc) + \mathcal{M}^*(\outvc \rightarrow \invc) = - \sum_N\int \mathcal{D\sigma}_N \mathcal{M}(\outvc \rightarrow \sigma) \mathcal{M}^*(\invc \rightarrow \sigma) (2\pi)^4\delta^{(4)}(p_{\outvc}+p_{\sigma}) \delta^{(4)}(p_{\invc}+p_{\sigma})
\label{krsunitarity21}	
\end{equation}
Now, we restrict this to the case when the incoming and outgoing states have 2 particles each (i.e., in the case of Compton amplitude). The intermediate states can have any number of particles
\begin{multline}
    \mathcal{M}(2_{\invc} \rightarrow 2_{\outvc}) + \mathcal{M}^*(2_{\outvc} \rightarrow 2_{\invc}) \\
	= -\sum_{N} \int \mathcal{D\sigma}_N \mathcal{M}(2_{\outvc} \rightarrow \sigma_n) \mathcal{M}^*(\invc \rightarrow \sigma_n) (2\pi)^4\delta^{(4)}(p_{\outvc}+p_{\sigma}) \delta^{(4)}(p_{\invc}+p_{\sigma})
\label{krsunitarity21}	
\end{multline}
Now, we incorporate the perturbative expansion \eqref{krsunitarity9} into account. 
\begin{equation}
	\mathcal{M}(i\rightarrow f)=\sum_{n=0}^\infty(\gem)^n\,\mathcal{M}^{(n)}(i\rightarrow f)
\end{equation}
where $\mathcal{M}^{(n)}(i\rightarrow f)$ is defined in terms of $T^{(n)} $ defined in \eqref{krsunitarity9}
\begin{equation}
    \langle f | T^{(n)} | i \rangle =  (2\pi)^4 \delta^4(p_i + p_f) \mathcal{M}^{(n)}(i\rightarrow f)
\end{equation} 
In this work, we are restricting to tree-level Compton amplitude, which is second order in $\gem^2$. So the \eqref{krsunitarity11} becomes (with $n=2$)
\begin{multline}
    \mathcal{M}^{(2)}(2_{\invc} \rightarrow 2_{\outvc}) + \left(\mathcal{M}^{(2)}\right)^*(2_{\outvc} \rightarrow 2_{\invc}) \\
	= -\sum_{n=0}^{2}\sum_{N} \int \mathcal{D\sigma}_N \mathcal{M}^{(n)}(2_{\outvc} \rightarrow \sigma_N) (\mathcal{M}^{(2-n)})^*(2_\invc \rightarrow \sigma_N) (2\pi)^4\delta^{(4)}(p_{\outvc}+p_{\sigma}) \delta^{(4)}(p_{\invc}+p_{\sigma})
\label{krsunitarity21}	
\end{multline}
So the only possible choice for $n$ is $1$, and in that case, the only possible for $N$ is also $1$. So, in the unitarity of tree-level Compton amplitudes, only one-particle states contribute. 
\begin{multline}
    \mathcal{M}^{(2)}(2_{\invc} \rightarrow 2_{\outvc}) + \left(\mathcal{M}^{(2)}\right)^*(2_{\outvc} \rightarrow 2_{\invc}) \\
	= - \sum_{I}\mathcal{M}^{(1)}(2_{\outvc} \rightarrow I) (\mathcal{M}^{(1)})^*(2_\invc \rightarrow I) (2\pi)^4\delta^{(4)}(p_{\outvc}+p_{I}) \delta^{(4)}(p_{\invc}+p_{I})
\label{krsunitarity21}	
\end{multline}
where $I$ is a one-particle state; the sum is over all one-particle internal states. So, the right-hand side is essentially a product of two three-point functions, and we need to sum over all possible one-particle states for the internal particle. 

So, we need to show that the left-hand side becomes a product of two three-point functions, where we need to sum up all possible one-particle states.

\subsection{Right-hand side}
\label{subsec:krsscomptonIunitarityII}

Let's first consider the right-hand side. The RHS is a product of two three-point functions, and we need to sum over all the states of the internal particles. The single-particle states are denoted by polarization 
\begin{equation}
	\zeta^{\mu_1\cdots\mu_\spin}({p; h})
\label{krsunitarity41}	
\end{equation}
$h$ is called the helicity(more details can be found in appendix \ref{app:completenessrelationproof}); it keeps track of the "spin" degrees of freedom. For example, in $3+1$ for spin-$\spin$ particles $h$ takes $2\spin+1$ values. Then the RHS (for $t$-channel) in \eqref{krsunitarity21} is given by 
\begin{equation}
	\sum_{h=-\spin}^{\spin}\frac{\epsilon_{4}\cdot k_1}{m}\bigg(\zeta_1\cdot \bar \zeta_I(k_1+k_4;h)\bigg)\bigg(\zeta_I(k_1+k_4; h)\cdot \bar \zeta_2\bigg)\frac{\epsilon_{3}\cdot k_2}{m}
\label{krsunitarity42}	
\end{equation}
$\zeta_I(k_1+k_4;h)$ is the polarization for the internal on-shell state. 
 
\subsubsection{Completeness relation for bosonic spin $\spin $ particle}
\label{subsubsec:krsscomptonIunitarityIII}

For the higher spin states one can show the following completeness relation(proved in appendix \eqref{app:completenessrelationproof})
\begin{equation}
	\sum_h \bar\zeta^{\mu_1\cdots\mu_\spin}(p;h) \zeta^{\nu_1\cdots\nu_\spin}(p;h)
	=\mathcal{P}^{(\spin)}_{(\mu),(\nu)}(p;\xi=\infty, m^2 )
\label{krsunitarity61}	
\end{equation}
Putting \eqref{krsunitarity61} in \eqref{krsunitarity42},  we get  
\begin{equation}
	(\varepsilon_4\cdot k_1)(\varepsilon_3\cdot k_2) \zeta_1^{\{\mu\}} \mathcal{P}^{(\spin)}_{\{\mu\},\{\nu\}}(k_1+k_4, \xi=\infty)\bar \zeta_2^{\{\nu\}} 
\label{krsunitarity62}	
\end{equation}
As long as we show that the LHS also gives the same quantity, we have proven that the scattering amplitude is unitary.

\subsection{Left-hand side}
\label{subsec:krsscomptonIunitarityIV}

Note that in \eqref{krsunitarity21}, on LHS, we have twice the real part of $\widetilde{\mathcal{M}}$. The amplitude looks purely imaginary as there is an overall $\iimg$ in all exchange amplitude. The origin of any kind of real part of the amplitude($\mathcal{M}$) is a consequence of using $\iimg \varepsilon$ in the propagator pole. So whatever analysis for the real part we are doing, we would only need the part which has poles, i.e. the exchange amplitudes($\widetilde{\mathcal{M}}$).  

We will first write some identities, which will be used in further calculations. Then, we would prove unitarity for $\xi=\infty$ since this case is the easiest; in the next one, we would show that for unitarity at $\xi=1$, we need to consider ghost exchanges with specific ghost couplings. Finally, we would show it is not possible to prove unitarity for any other $\xi$.

Before we begin, let us go through an identity which is at the heart of all the following calculations. It is given by,
\begin{equation}
    \label{eq:im_of_pole}
    \lim_{\varepsilon\rightarrow0}\mathrm{Im}\Bigg[\frac{1}{x-\iimg \varepsilon}\Bigg] = \lim_{\varepsilon\rightarrow0}\frac{1}{2\iimg}\Bigg[\frac{1}{x-\iimg \varepsilon} - \frac{1}{x+\iimg \varepsilon}\Bigg] = \lim_{\varepsilon\rightarrow0} \frac{\varepsilon}{x^2+\varepsilon^2}=\pi\delta(x)
\end{equation}
The appearance of $\delta(x)$ should be obvious. A factor of $\pi$ is multiplied because $\int_{-\infty}^{\infty} \frac{\varepsilon dx}{x^2+\varepsilon^2} = \pi$.

We can apply this to the propagator; it has factors like $-\iimg/(k^2+m^2-\iimg \epsilon)$,   $-\iimg/(k^2+\xi m^2-\iimg \epsilon)$. From the above identity, we can see that every simple pole contributes to the real part; it puts a delta function, which essentially puts an on-shell condition for the internal leg. 

Let's start with the scalar propagator
\begin{equation}
	\frac{-\iimg }{k^2+m^2-\iimg \epsilon}
\label{krsunitarity201}	
\end{equation}
This implies that the real part of the pole of the propagator is a delta function for the internal legs. The higher spin propagator is given by 
\begin{eqnarray}
		\frac{-\iimg }{k^2+\xi m^2-\iimg \epsilon}	\mathcal{P}^{(\spin)}_{(\mu),(\nu)}(k;\xi, m^2 )
\label{krsunitarity202}	
\end{eqnarray}
This has poles at $m^2$ and $\xi m^2$. We start with the pole at $m^2$. In that case, one can check that 
\begin{equation}
	\textrm{Residue}\left(\mathcal{P}^{(\spin)}_{(\mu),(\nu)}(k;\xi, m^2 )\right)\Big|_{k^2=-m^2,\xi\ne 1}=\textrm{Residue}\left(\mathcal{P}^{(\spin)}_{(\mu),(\nu)}(k;\infty, m^2 )\right)\Big|_{k^2=-m^2}
\label{krsunitarity203}	
\end{equation}
A propagator at general $\xi$ has more poles at $k^2=-\xi m^2$, which can also contribute to the real part. So, let's consider the two simple cases first: $\xi=\infty$ and $\xi=1$. In both cases, the propagator has only one pole.

\subsubsection{$\xi=\infty$}
\label{subsubsec:krsscomptonIunitarityV}

In the case of $\xi=\infty$, the numerator of the propagator is precisely the RHS of \eqref{krsunitarity203}. So, if we consider the amplitude for $\xi=\infty$ and compute the residue, it precisely becomes  \eqref{krsunitarity62}. 

\subsubsection{$\xi$ finite}
\label{subsubsec:krsscomptonIunitarityVII}
The computations are more involved for general $\xi$. In fact, we will show that the amplitude is not unitarity for general $\xi$; only $\xi=1$ and $\xi=\infty$ works. First, note that $\xi=\infty$ is unitary, as we proved in the previous subsection. So whatever amplitude we get for other $\xi$s, as discussed earlier, the non-analytic piece of the difference should come from unphysical exchanges (a.k.a auxiliary field or ghost exchanges). Such unphysical exchanges are required even in the case of Standard Model to ensure unitarity in $R_\xi$ gauge\cite{Weinberg:1996kr}. 

The exchange amplitude for $\xi=\infty$ is given in the \eqref{comptonminimalcalculation42} and for general $\xi$ in \eqref{comptonminimalcalculation26}. The difference is,
\begin{equation}
    -\iimg \frac{\gem^2}{m^2}\frac{\spp{\varepsilon_3}{k_2}\spp{\varepsilon_4}{k_1}}{t+m^2-\iimg \epsilon}\sum\limits_{b=0}^{\spin-1} B(\spin, b, D) \spp{\zeta_1}{\zeta_2}^b \spp{\zeta_1}{k_4}^{\spin-b}\spp{\zeta_2}{k_3}^{\spin-b}\bigg[\left(\frac{1}{m^2}\right)^{\spin-b} -\ \left(\frac{\xi-1}{t+\xi m^2-\iimg \varepsilon}\right)^{\spin-b}\bigg]
\end{equation}
Since the $\xi=\infty$ exchange amplitude is already unitary, this difference of the general $\xi$ and $\xi=\infty$ amplitude should be compensated by unphysical exchanges. To find these we need to write the above difference as a sum over varying order in poles. In the above expression, consider taking the pole sitting outside the summation inside the square bracket. Now using the identity,
\begin{equation}
    \label{identity2}
    \frac{1}{t+m^2-\iimg \epsilon} \bigg(\frac{\xi-1}{t+\xi m^2-\iimg \epsilon}\bigg)^n = \frac{1}{t+m^2-\iimg \epsilon}\bigg(\frac{1}{m^2}\bigg)^n - \sum\limits_{i=0}^{n-1} \frac{(\xi-1)^i}{(t+\xi m^2 -\iimg \epsilon)^{i+1}}\bigg(\frac{1}{m^2}\bigg)^{n-i}
\end{equation}
we get the expression as,
\begin{equation}
\label{eq:xiinf-xi1}
    -\iimg \frac{\gem^2}{m^2}\spp{\varepsilon_3}{k_2}\spp{\varepsilon_4}{k_1}\sum\limits_{b=0}^{\spin-1} B(\spin, b, D) \spp{\zeta_1}{\zeta_2}^b \spp{\zeta_1}{k_4}^{\spin-b}\spp{\zeta_2}{k_3}^{\spin-b} \sum\limits_{i=0}^{\spin-b-1} \frac{(\xi-1)^i}{(t+\xi m^2 -\iimg \epsilon)^{i+1}}\bigg(\frac{1}{m^2}\bigg)^{\spin-b-i}
\end{equation}

Note that the $m^2$ pole is not present anymore; instead, only the $\xi m^2$ are left. This means the fields of unphysical exchanges need the mass to be $\sqrt{\xi}m$. Moreover, the terms that the terms present have poles of maximum degree $\spin$ and minimum degree $1$. Since these poles can only come from the propagator of unphysical fields, which could be constructed just like the case of Massive Higher Spin particles in \ref{subsec:krsscomptonIbhsreview}, the unphysical fields have to be of the Spin-$0$ till Spin-$(\spin-1)$. We need to find a vertex which will make the amplitude unitary for general $\xi$.

As a stepping stone, let us first do it for $\spin=1$. From \eqref{eq:xiinf-xi1}, we get,
\begin{equation}
     -\iimg \frac{\gem^2}{m^4}\spp{\varepsilon_3}{k_2}\spp{\varepsilon_4}{k_1}\frac{\spp{\zeta_1}{k_4}\spp{\zeta_2}{k_3}}{(t+\xi m^2 - \iimg \epsilon)}
\end{equation}
It is very easy to see vertex and unphysical field propagators needed,
\begin{equation}
\label{s0_ghost_vertex_and_prop}
    \pm\iimg\frac{\gem}{m^2}\spp{\zeta_1}{k_3} \spp{\varepsilon_3}{k_{12}} \qquad\qquad\qquad  \frac{-\iimg}{t+ \xi m^2 -\iimg \epsilon}
\end{equation}
This case is parallel to the standard model. The first unknown territory is that of $\spin=2$. We will do that next.

 For $\spin=2$, \eqref{eq:xiinf-xi1} becomes,
 \begin{align}
  \label{eq:spin2diff}
  &-\iimg \frac{\gem^2}{m^4}\spp{\varepsilon_3}{k_2}\spp{\varepsilon_4}{k_1}\spp{\zeta_1}{k_4}\spp{\zeta_2}{k_3}\nonumber\\
  &\times \Bigg[ \frac{B(2, 0, D)}{m^2} \spp{\zeta_1}{k_4}\spp{\zeta_2}{k_3}\bigg\{\frac{1}{(t+\xi m^2 -\iimg \epsilon)} + \frac{(\xi-1)m^2}{(t+\xi m^2 -\iimg \epsilon)^2}\bigg\} 
   + B(2, 1, D) \spp{\zeta_1}{\zeta_2} \frac{1}{(t+\xi m^2 -\iimg\epsilon)}
  \Bigg]
\end{align}
these have to be compensated by Spin-$0$ and Spin-$1$, unphysical exchanges. The possible vertices for spin-$1$ unphysical field are,
\begin{align}
    &\frac{\lambda_1}{2m^2} \spp{\zeta_1}{k_3}\spp{\zeta_1}{\phi_2}\spp{\varepsilon_3}{k_{12}}\label{eq:lambda1_vert}\\
    &\frac{\lambda_2}{2m^4} \spp{\zeta_1}{k_3}^2\spp{\phi_2}{k_3}\spp{\varepsilon_3}{k_{12}}
\end{align}
and for spin-$0$ unphysical fields are,
\begin{equation}
    \frac{\lambda_0}{2m^3} \spp{\zeta_1}{k_3}^2\spp{\varepsilon_3}{k_{12}}
\end{equation}
Amplitude for the spin-$1$ unphysical particles for the above vertices is given by,
\begin{align}
	\label{eq:spin21ghostamp}
	\frac{\iimg}{m^4}\spp{\varepsilon_3}{k_2}&\spp{\varepsilon_4}{k_1}\spp{\zeta_1}{k_4}\spp{\zeta_2}{k_3}
	\Bigg[
		\lambda _1{}^2 \bigg\{\frac{\spp{\zeta _1}{\zeta _2}}{t+\xi m^2}-\frac{(\xi-1) \spp{k_4}{\zeta _1} \spp{k_3}{\zeta _2}}{(t+\xi m^2)^2}\bigg\}\nonumber\\
		&+\frac{\lambda _1 \lambda _2}{m^2} \bigg\{-\frac{(\xi -1) (m^2+t) \spp{k_4}{\zeta _1} \spp{k_3}{\zeta _2}}{(t+\xi m^2)^2}+\frac{\spp{k_3}{\zeta _1} \spp{k_3}{\zeta _2}}{t+\xi m^2}+\frac{\spp{k_4}{\zeta _1} \spp{k_4}{\zeta _2}}{t+\xi m^2}\bigg\}\nonumber\\
		&+\frac{\lambda _2{}^2 (m^4 (1-5 \xi )-2 m^2 (2 \xi  t+t+\xi  u)-t (\xi  t+t+2 u)) \spp{k_4}{\zeta _1} \spp{k_3}{\zeta _2}}{4 m^4 (t+\xi m^2)^2}
		\Bigg]
\end{align}
Keeping in mind that this amplitude should be subtracted from \eqref{eq:spin2diff}, note the following observations,
\begin{itemize}
	\item The last term of \eqref{eq:spin2diff} is generated in the above amplitude with only $\lambda_1^2$ coefficient. hence
	\begin{equation}
		\lambda_1 = \pm\iimg\gem\sqrt{B(2,1,D)}
	\end{equation} 
	\item there are no terms like $\spp{k_4}{\zeta _1} \spp{k_4}{\zeta _2}$ or $\spp{k_3}{\zeta _1} \spp{k_3}{\zeta _2}$ in the \eqref{eq:spin2diff} but these terms appear in the above amplitude with $\lambda_1 \lambda_2$ as coefficient. hence
	\begin{equation}
		\lambda_2 = 0
	\end{equation}
\end{itemize}
After putting these values in \eqref{eq:spin21ghostamp} and subtracting it from \eqref{eq:spin2diff}, we get,
\begin{align}
	\label{eq:spin2diffv2}
	&-\iimg \frac{\gem^2}{m^4}\spp{\varepsilon_3}{k_2}\spp{\varepsilon_4}{k_1}\spp{\zeta_1}{k_4}\spp{\zeta_2}{k_3}\nonumber\\
	&\times \Bigg[ \frac{B(2, 0, D)}{m^2} \spp{\zeta_1}{k_4}\spp{\zeta_2}{k_3}\bigg\{\frac{1}{(t+\xi m^2 -\iimg \epsilon)} + \frac{(1+B(2,1,D))(\xi-1)m^2}{(t+\xi m^2 -\iimg \epsilon)^2}\bigg\}
	\Bigg]
\end{align}
We are still left with second order pole in $\xi m^2$ which can not be removed by a scalar(spin-$0$) unphysical particle exchange. If we set $\xi=1$, then the second order pole term drops out, we can simply set 
\begin{equation}
	\lambda_0 = \pm \iimg \gem \sqrt{B(2,0,D)}
\end{equation}
and subtract the spin-$0$ exchange contribution from \eqref{eq:spin2diffv2} to make it zero. 

So finally the result is that although for spin-$1$, we could show unitarity of the amplitude in general $\xi$ by adding appropriate unphysical exchange, we could not do the same for spin-$2$. For this, only $\xi=1$ amplitude can be made unitary by adding the appropriate unphysical exchanges. These arguments can be extended for any arbitrary integer spin. Hence it is not possible to have a unitary amplitude for arbitrary spin in general $\xi$ gauge, only $\xi=1$ and $\xi=\infty$ works. 

Now that we know which $\xi$ value could give us unitary amplitude, let us determine the three point vertex involving unphysical particle required to make the amplitude unitary for arbitrary spin. Putting $\xi=1$ in \eqref{eq:xiinf-xi1}, only the $i=0$ term in the inner sum remains, rest are zero. We are simply left with single order pole in $\xi m^2$. It is given by,
\begin{equation}
	-\iimg \frac{\gem^2}{m^2}\spp{\varepsilon_3}{k_2}\spp{\varepsilon_4}{k_1}\sum\limits_{b=0}^{\spin-1} \frac{B(\spin, b, D)}{m^{2(S-b)}(t+\xi m^2-\iimg\epsilon)} \spp{\zeta_1}{\zeta_2}^b \spp{\zeta_1}{k_4}^{\spin-b}\spp{\zeta_2}{k_3}^{\spin-b}
\end{equation}
At $\xi=1$, our proposed propagator for unphysical field of spin-s is simply,
\begin{equation}
	\frac{-\iimg\eta^{\mu_1(\nu_1}\eta^{\mu_2\nu_2}\dots\eta^{\mu_s\nu_s)}}{t+\xi m^2-\iimg\epsilon}
\end{equation}
The $\eta$s simply contract the unphysical polarisation stripped vertex from the left and right. To get the $\spp{\zeta_1}{\zeta_2}^b$ we need the vertex to be spin-b unphysical polarisation to be contracted with $b$ numbers of $\zeta$s both, on left and right so that after stripping the them glueing using the above propagator we get the desired term $\spp{\zeta_1}{\zeta_2}^b$. Rest $\zeta$s on both sides be contracted with $k_3$ or $k_4$ depending on left or right vertex. So the vertex is given like below,
\begin{equation}
	\label{ghostcoupling}
    \pm\iimg\frac{\gem\sqrt{B(\spin,b,D)}}{m^{\spin-b+1}}\spp{\zeta_1}{\phi_2}^b  \spp{\zeta_1}{k_3}^{\spin-b} \spp{\varepsilon}{k_{12}}
\end{equation} 

We can get the position space vertex which can be added to a lagrangain(once we have one) directly from the expression of ghost vertex in \eqref{ghostcoupling} by replacing each momenta with $-\iimg\partial_{\_}$. But the lagrangian term should be real, hence depending on if there are even or odd momentas, which is determined also by the spin of the ghost, there should be an $\iimg$ or not in the above expression. To verify this, we note that sign of $B(\spin, b, D)$ is determined completely by $(-1)^{\spin-b}$ because the rest is completely positive. $\sqrt{B(\spin, b, D)}$ gives $\iimg^{\spin-b}$, there is already an $\iimg$ sitting outside, which finally gives $\iimg^{\spin-b+1}$. In the above expression\eqref{ghostcoupling} as well, we have $\spin-b+1$ momentas. So this vertex is good enough to give a real interaction vertex in the lagrangian finally. The lagrangian term for the above vertex will be given by,
\begin{align}
	\frac{\gem}{m^{\spin-b+1}} &\bigg(\sum_{a=0}^{\lfloor\frac{\spin-b}{2}\rfloor} {}^{\spin-2a}\mathrm{C}_{b} \sqrt{A(\spin,a,D)}\bigg)\nonumber\\
	& (\partial_\rho\Phi^{\mu_1\mu_2\dots\mu_\spin} \bar\varphi_{\mu_1\mu_2\dots\mu_b}-\Phi^{\mu_1\mu_2\dots\mu_\spin} \partial_\rho\bar\varphi_{\mu_1\mu_2\dots\mu_b})\partial_{\mu_{b+1}}\partial_{\mu_{b+2}}\dots\partial_{\mu_{\spin}} A^\rho + \mathrm{h.c.}
\end{align}
h.c. denotes hermitian conjugate of the previous term.

\section{Conclusion and future directions}
\label{sec:krsscomptonIconclusion}

The search to find the rules of a perturbative quantum field theory with higher spin particles is still ongoing. In this work, we have computed the tree-level four-point electromagnetic Compton amplitude. We did it for bosonic higher spin particles, which are completely symmetric, traceless tensors of rank $\spin$; in $3+1$ dimensions, these particles have spin-$\spin$. We found that the contributions from the exchange diagrams are not gauge invariant; one needs to add four-point contact terms to make it unitary. We provided the algorithm to construct the contact terms which can be applied to higher points as well. We also introduced an analogue of $R_\xi$ gauge. The propagator that is commonly found in the literature corresponds to $\xi=0$. In this work, we have shown that for any finite value of $\xi\ne1$, it is not possible to have a unitary theory. For $\xi=\infty$, the Compton amplitude can be Unitary just by adding contact terms. For $\xi=1$, it is also possible to have a unitary theory, but at the cost of adding a tower of unphysical particles with spin ranging from $0$ to $\spin-1$. To summarize, it is possible to have a unitary theory only for two values of $\xi$: $1$ and $\infty$.

$\xi=\infty$ is known as unitary gauge. It is useful since only physical particles propagate in this gauge. However, in case of Higgsed theory the loop amplitudes are badly divergent; in fact, the renormalizability of the Standard Model is still missing for the $\xi=\infty$ gauge. On the other hand, $\xi=1$, the unphysical goldstone boson, has the same mass as that of the gauge bosons and for Higgsed theory it is possible to show that theory is renormalizable. In fact, the string theory spectrum is also available only in the $\xi=1$ gauge. It would be interesting to compute loop amplitudes in the higher spin theories to figure out whether $\xi=1$ and $\xi=\infty$ are equally good for loop amplitudes or not. 

Tree-level higher point scattering amplitudes are also interesting and important. In particular, one would like to compute $n-2$ massless, 2 massive scattering amplitude. From this scattering amplitude, one can use unitarity method to construct higher loop amplitudes. In scalar QED, the four point contact term is enough to ensure Unitarity at the higher point functions; however, for Higgsed massive spin-$1$ QED, one needs to incorporate two photon, two ghost vertices to ensure unitarity at higher point. It would interesting to check what happens for higher point function in case of higher spin particles: whether four point contact terms are enough or one needs to include higher point contact terms like Einstein's gravity. Another important consistency check is to compute loop corrections. In order to compute loop amplitudes, we need four-point contact interactions with two massive bosons and two unphysical particles. It is possible to find out such interactions by looking at the unitarity of 5/6 point tree-level scattering. Thus tree level higher point scattering is important to before proceeding to understand loop corrections.

In this work, we focused on the Bosonic higher spin particles, which are completely symmetric and traceless. In spacetimes with dimensions $>4$, there are also mixed symmetric boson particles. One can also consider fermionic higher spin particles interacting with photons. It would be interesting to check whether the features of the Compton amplitude, found in this work, remain the same for mixed symmetric higher spin particles and for fermionic particles \footnote{We are thankful to Debapriyo Chowdhury for comments on this point.}.

Even if we restrict to the case of symmetric traceless bosonic higher spin particles, there are $2\spin+1$ kinematically distinct three-point vertices involving two higher spin particles with $\spin$ and photon. In this work, we restrict to the interaction that dominates in the low energy/long distance. For a complete understanding of the Compton amplitudes, all these three-point interactions are important; for example, the high energy behaviour of the Compton amplitude is also controlled by the other three-point functions. These kinematically distinct three-point functions are different multipole moments of the quantum particles. The following one is related to the magnetic moment of the quantum particles
\begin{equation}
	(-\iimg\,)\frac{\gcou_1}{m}(\zeta_1\cdot \bar\zeta_2)^{\spin -1} \Big(-(\zeta_1\cdot \varepsilon_3)	(\bar\zeta_2\cdot k_{3}) 
+(\bar\zeta_2\cdot \varepsilon_3)( \zeta_1\cdot k_{3})	\Big)
\end{equation}
$\gcou_1$ is called anomalous magnetic moment. An interesting issue related to this is the value of anomalous magnetic moments for higher spin particles. It is believed that the anomalous magnetic moment is 2 for all higher spin particles. In the context of string theory, this was shown in \cite{Argyres:1989cu}. Weinberg argued the value of anomalous magnetic moment to be two from the perspective of better ultraviolet behaviour \cite{Weinberg:1970etn}. One can also ask whether there is any natural value for the higher moments (quadrupole, octupole ...) and whether it is possible to find those values from the consideration of higher energy behaviour of Compton amplitude of higher spin particles. For example, in string theory, all these couplings appear with a very particular co-efficient \cite{Sagnotti:2010at}. From the perspective of EFT, it is not clear whether string theory values appear at any special points on this $2\spin+1$ dimensional space of coupling constants. Thus, computing Compton amplitude for most general three-point functions is necessary to answer some of these questions. In the current analysis, we could find gauge invariant answer only for $\xi=1$ and $\infty$. Once we have the answer for the most general three point function one can check whether it is possible to make the answer $\xi$ independent or not at some special points in this space of three-point functions. The key challenge in those computations is the huge form of propagator and the non symmetric higher spin vertices. However, an extension of the derivative operator introduced in \cite{Balasubramanian:2021act} can help us in answering those questions. 

In 3+1 dimensional scattering, spinor helicity variables \cite{Elvang:2015rqa} have turned out to be an efficient tool. The formalism to compute scattering	 amplitudes with massive particles is also available \cite{Arkani-Hamed:2017jhn}.  In recent times, there has been a lot of interest in computing gravitational Compton amplitude and mostly the spinor helicity variables have been used to address those problems\cite{Arkani-Hamed:2017jhn, Arkani-Hamed:2019ymq}. The gravitational compute amplitude is crucial to know about the scattering of gravitational radiation from spinning black holes. We would like to extend our computation to the spinor helicity variables and also to the Yang-Mills and gravitational Compton amplitude. The black holes can be modelled as quantum particles with very precise values for the multipole moments \cite{Arkani-Hamed:2019ymq}. The values of the multipole moments for other astrophysical objects like neutron stars are different from the black holes.  Thus, understanding gravitational Compton amplitude for most general three-points can become an essential tool in gravitational wave spectroscopy.

On the formal side, in this work, we have taken an on-shell approach. \footnote{We thank Ashoke Sen for this suggestion. We thank Sandip Trivedi, Prahar Mitra for questions on this point during NSM.}It would be good to write down Lagrangian and find the Feynman rules from the Lagrangian. One could use string field theory \cite{Sen:2024nfd} to find the corresponding action\footnote{We thank Mritunjay Kumar Verma, Spenta Wadia and especially Ashoke Sen for discussions on this point.} and check the results in this work. Finally, the existence of Lagrangian is important for understanding non-perturbative quantum field theory; on-shell tools are enough to address those questions. 

We hope to answer some of these questions in the near future. 

\paragraph{Acknowledgement} We are thankful to R Loganayagam and Shridhar Vinayak for many discussions on higher spin particles. AR would like to thank Arindam Tarafdar for many discussions on this topic and for a collaboration at the early stage of the project; this was reported in a \href{https://drive.google.com/file/d/1DlcQqyjMC3Mgv1Je0gFCtuQdgSZuAr5y/view?usp=sharing}{MS thesis} at IISER Bhopal. We thank Snehasis Das for many comments on the first version of the draft. We are thankful to Mahesh Balasubramanian, Sanjoy Biswas, Debapriyo Chowdhury, Snehasis Das, Godwin Martin, Shibashis Mukhopadhyay, Raj Patil and Mritunjay Verma for their comment on a preliminary version of this report. AK would like to acknowledge the support from IISER Bhopal through the fellowship for PhD students. RS gratefully acknowledges support from CSIR. We thank the members of \href{https://sites.google.com/iiserb.ac.in/iiserbstrings/home}{strings@iiserb}, Department of Physics, IISERB, for providing a vibrant atmosphere. A preliminary version of this work was presented at \href{https://iitrpr.ac.in/nsm/}{National strings meeting} at IIT Ropar and at \href{https://strings19.wixsite.com/centre-for-theoretic/about-3}{amplitudes@iiti} at IIT Indore. AR would like to acknowledge the hospitality of ICTS, Bangalore towards the completion of the work; we have received many scientific input during the \href{https://www.youtube.com/watch?v=ZfUnEu1IyUQ}{string seminar at ICTS}. 

Finally, we are grateful to the people of India for their generous support for research in basic sciences.

\newpage
\appendix
\section{Notation and convention}
\label{app:krssInotation}

\begin{subequations}   
\begin{eqnarray} 
\textrm{Spacetime metric}\quad \quad && \eta_{\mu \nu }=\diag(-1,1,\cdots,1)
\\
\textrm{Index for massless particle/Null vector}\quad \quad &&  i,j
\\
\textrm{Index for massive particle/Time-like vector}\quad \quad && a,b 
\\
\textrm{Lorentz indices}\quad \quad && \mu,\nu 
\\
\textrm{Momentum}\quad \quad && k_\mu 
\\
\textrm{Mandestam variables}\quad \quad && s,t,u
\\
\textrm{Spin}\quad \quad && \spin 
\\
\textrm{Helicity}\quad \quad && h
\\
\textrm{Polarization of a massless particle}\quad \quad && \epsilon_\mu 
\\
\textrm{Polarization of a massive particle}\quad \quad && \zeta_\mu 
\\
\textrm{Mass parameters} \quad\quad && m,m_a
\\
\textrm{Spinor Helicity $x$ factor} \quad\quad && x, x_{ia}(h_i)
\\
\textrm{Coupling contants} \quad\quad && e
\nonumber\\
\textrm{Momentum difference}\quad \quad && k_{ab}=k_a-k_b
\\
\textrm{Linearized Maxwell field strength}\quad \quad && \mathcal{B}_{\mu \nu}
\end{eqnarray} 
\end{subequations} 
We follow the following convention for the Mandelstam variables 
\begin{eqnarray}
s=-(k_1+k_2)^2
&\quad,\quad &
t=-(k_1+k_4)^2\quad \text{and}
\nonumber\\
&u=-(k_1+k_3)^2&~.
\label{hscomp22} 
\end{eqnarray}
This is the same as a convention in Green-Schwarz-Witten \cite{Green:2012oqa} \footnote{vol.1 page 373, 378} but different from Polchinski \cite{Polchinski:1998rq}. We also follow the convention such that all the {\it external particles} are {\it outgoing}.

\section{Completeness relation for genaral dimension any integer spin}
\label{app:completenessrelationproof}
Any local quantum field can be expanded into its modes as,
\begin{equation}
	\Phi^{\mu_1\mu_2\dots\mu_\spin}(x) = \int \frac{d^{D-1}p}{(2\pi)^3}\frac{1}{2E}\bigg\{\sum_{h}\zeta^{\mu_1\mu_2\dots\mu_\spin}(p; h)a_{\vec{p},h}e^{\iimg p\cdot x}+\bar\zeta^{\mu_1\mu_2\dots\mu_\spin}(p; h)a^\dagger_{\vec{p},h}e^{-\iimg p\cdot x}\bigg\}
\end{equation}
where $h$ denotes all the quantum numbers except the spacetime translation symmetry quantum number $p$. We denote the momenta with $p$ and momentum by $\vec{p}$. $h$ can be understood as collection of eigenvalues of the cartans of the little group, $SO(D-1)$. For example, for the dimension,$D=3+1$, There is only one cartan number whose value ranges from $-\spin,\dots,0,\dots,\spin$ for spin-$\spin$, hence $h=-\spin,\dots,0,\dots,\spin$. For higher dimension, $h$ will consist of more than one numbers.  $\zeta$s are called polarisation and they are such that, they follow following relation,
\begin{equation}
	\sum_{h'} \mathcal{D}(W(\Lambda, p))_{h h'} \zeta^{\mu_1\mu_2\dots\mu_\spin}(\Lambda p; h') = \sum_{\nu_1\nu_2\dots\nu_\spin} \Lambda^{\mu_1}{}_{\nu_1}\Lambda^{\mu_2}{}_{\nu_2}\dots\Lambda^{\mu_\spin}{}_{\nu_\spin}\zeta^{\nu_1\nu_2\dots\nu_\spin}(p; h)
\end{equation}
where $W(\Lambda, p)=L^{-1}(\Lambda p)\Lambda L(p)$ are the little group representation. $\mathcal{D}$ is the little group representation matrix. The boost transformation, $L(q)$ boosts the momenta from standard momenta $(m, \vec{0})$ to $(q_0, \vec{q})$, given by,
\begin{align}
	\label{eq:boostmatrix}
	L^0{}_0(q) &= \gamma\\
	L^i{}_k(q) &= \delta_{ik}+(\gamma-1)\hat{q}_i\hat{q}_k\\
	L^i{}_0(q) &= L^0{}_i(q) = \hat{q}_i\sqrt{\gamma^2-1}
\end{align}
where
\begin{equation}
	\hat{q}_i = \frac{q_i}{|\vec{q}|}\qquad \mathrm{and}\qquad \gamma = \frac{\sqrt{|\vec{q}|^2+m^2}}{m}
\end{equation}
details of this can be found in chapter 2 of \cite{Weinberg:1995mt}.

We can get the polarisation for any frame by boosting it using these operators. In the rest frame of the massive spin-$\spin$ particle, due transversality, component of the polarisation tensor along time direction is vanishing, i.e.,

\begin{equation}
	\zeta^{\dots0\dots}(m, \vec{0}; h)= 0
\end{equation}

Hence, we can denote the polarisations indices with $i,j=1\dots D-1$. Now consider the outer product of these polarisarions,
\begin{equation}
	T^{i_1 i_2\dots i_\spin,j_1 j_2\dots j_\spin}(m, \vec{0}) \equiv \sum_{h} (\zeta^{i_1 i_2\dots i_\spin}(m, \vec{0}; h))^*\zeta^{j_1 j_2\dots j_\spin}(m, \vec{0}; h)
\end{equation}
Noting that polarisations are symmetric traceless, and the only tensor available to us in this just space is $\delta^{ij}$, we can write the most general ansatz for $T$ as,
\begin{equation}
	T^{i_1 i_2\dots i_\spin,j_1 j_2\dots j_\spin}(m, \vec{0}) = \sum_{a=0}^{\lfloor\frac{\spin}{2}\rfloor} A(\spin, a, D)\bigg[\delta^{i_1 i_2}\delta^{j_1 j_2}\dots\delta^{i_{2a-1} i_{2a}}\delta^{j_{2a-1} j_{2a}}\delta^{i_{2a+1}j_{2a+1}}\dots\delta^{i_\spin j_\spin}\bigg]_{\mathrm{sym}(i,j)}
\end{equation}
Then we demand tracelessness out of the above ansatz, this fixes the $A(\spin,a,D)$ to \eqref{raamhspinreview11}. Now, to get the value of $T$ in any other frame, let us boost this. 
\begin{equation}
	\label{Tmunu}
	T^{\mu_1 \mu_2\dots \mu_\spin,\nu_1 \nu_2\dots \nu_\spin}(p) = L^{\mu_1}{}_{i_1}(p) L^{\mu_2}{}_{i_2}(p) \dots L^{\mu_\spin}{}_{i_\spin}(p) L^{\nu_1}{}_{j_1}(p) L^{\nu_2}{}_{j_2}(p) \dots L^{\nu_\spin}{}_{j_\spin}(p) T^{i_1 i_2\dots i_\spin,j_1 j_2\dots j_\spin}(m, \vec{0})
\end{equation}
Using the expressions from \eqref{eq:boostmatrix}, one can check that,
\begin{align}
	L^0{}_i L^0{}_j \delta^{ij} &=  -1 + \frac{p^0 p^0}{m^2}\\
	L^0{}_i L^k{}_j \delta^{ij} &=  p^0 p^k\\
	L^k{}_i L^l{}_j \delta_{ij} &= \delta^{kl}+\frac{p^k p^l}{m^2}
\end{align}
which can be compressed into one line as,
\begin{equation}
	L^\mu{}_i L^\nu{}_j \delta_{ij} = \eta^{\mu\nu}+\frac{p^\mu p^\nu}{m^2} \equiv \Theta^{\mu\nu}(p, \infty)
\end{equation}
Note that in the RHS of \eqref{Tmunu}, this will used for all $LL\delta$ contraction. Finally, the completeness relation is given as,
\begin{align}
	T^{\mu_1 \mu_2\dots \mu_\spin,\nu_1 \nu_2\dots \nu_\spin}(p) = \Bigg[\sum_{a=0}^{\lfloor \spin/2\rfloor}
    A(\spin,a,D)\, 
    \Theta_{\mu_1\mu_2}(p, \infty)\Theta_{\nu_1\nu_2}(p, \infty)&\dots\Theta_{\mu_{2a-1}\mu_{2a}}(p, \infty)\Theta_{\nu_{2a-1}\nu_{2a}}(p)\nonumber\\
	&\Theta_{\mu_{2a+1}\nu_{2a+1}}(p, \infty)
    \dots\Theta_{\mu_\spin\nu_\spin}(p, \infty)\Bigg]_{\text{sym} (\mu,\nu)} 
\end{align} 
where are $\Theta^{\mu\nu}$ is as defined in \eqref{comptoncalculation1}. Note that this is also the identity of the symmetric, traceless and transverse subspace of an rank-$\spin$ tensorial (vector) space. 

\section{Trace property of Projector}
\label{traceproperty}
Consider the following trace operation on Spin-$\spin$ projector defined in \eqref{raamhspinreview4.1}
\begin{equation}
	\eta^{\mu_\spin\nu_\spin}\mathcal{P}^{(\spin)}_{\mu_1\cdots \mu_\spin;\nu_1\cdots \nu_\spin}(p) 
\end{equation}
Since, it has to be symmetric traceless transverse rank-$2(\spin-1)$ tensor, it has to be proportional to projector of rank-${\spin-1}$ field which can be written as,
\begin{equation}
	\eta^{\mu_\spin\nu_\spin}\mathcal{P}^{(\spin)}_{(\mu), (\nu)}(p) = K_\spin \mathcal{P}^{(\spin-1)}_{(\mu), (\nu)}(p) 
\end{equation}
We are yet to find the proportianality constant $K_\spin$. Now consider full trace of the projector,
\begin{align}
	&\prod_{i=1}^{\spin}\eta^{\mu_i\nu_i}\mathcal{P}^{(\spin)}_{(\mu), (\nu)}(p) = K_\spin K_{\spin-1}\dots K_{1} \\
	\implies & K_\spin =\frac{\prod_{i=1}^{\spin}\eta^{\mu_i\nu_i}\mathcal{P}^{(\spin)}_{(\mu), (\nu)}(p)}{\prod_{i=1}^{\spin-1}\eta^{\mu_i\nu_i}\mathcal{P}^{(\spin-1)}_{(\mu), (\nu)}(p)}
\end{align}
Since the projector is the identity of symmetric traceless transverse subspace, its trace would give the dimension of the symmetric traceless transverse representation which is the degree of freedom. Hence, 
\begin{equation}
	\eta^{\mu_\spin\nu_\spin}\mathcal{P}^{(\spin)}_{(\mu), (\nu)}(p) = \frac{\mathrm{dof}(\spin, D)}{\mathrm{dof}(\spin-1, D)} \mathcal{P}^{(\spin-1)}_{(\mu), (\nu)}(p) 
\end{equation}
where dof$(\spin, D)$ denotes the degrees of freedom of an rank-$\spin$ symmetric traceless and transverse field in $D$ spacetime dimension.

\bibliographystyle{utphys}
\bibliography{emcomptonbiblio} 

\end{document}